\documentclass[%
 twocolumn,
 aps,pra,
 amsmath,amssymb,
 superscriptaddress,
 nofootinbib,
 reprint,%
]{revtex4-2}
\usepackage[utf8]{inputenc} 
\usepackage{stmaryrd}
\usepackage{color}
\usepackage{array}
\usepackage{enumitem}
\usepackage{mathpazo}
\usepackage{setspace}
\usepackage{bm}
\usepackage{amssymb}
\usepackage{amsthm}
\usepackage{mathtools}
\usepackage{multirow}
\usepackage{cases}
\usepackage[colorlinks=true,linkcolor=blue,citecolor=blue,pdfauthor={ },pdftitle={ },pdfsubject={ },pdfkeywords={ }]{hyperref}
\usepackage{pgfplots}
\usepackage{lipsum}
\usepackage{youngtab}
\usepackage{verbatim}
\usepackage[capitalise]{cleveref}
\usepackage{subcaption}
\usepackage{algorithm}
\usepackage{algpseudocode}
\usepackage{float}

\newtheorem{definition}{Definition}
\newtheorem{proposition}[definition]{Proposition}
\newtheorem{lemma}[definition]{Lemma}

\newtheorem{theorem}[definition]{Theorem}

\newtheorem{corollary}[definition]{Corollary}
\newtheorem{conjecture}[definition]{Conjecture}

\newtheorem{remark}[definition]{Remark}
\newtheorem{example}[definition]{Example}
\newtheorem{question}[definition]{Question}

\def\Dbar{\leavevmode\lower.6ex\hbox to 0pt
{\hskip-.23ex\accent"16\hss}D}
\makeatletter
\def\url@leostyle{%
  \@ifundefined{selectfont}{\def\UrlFont{\sf}}{\def\UrlFont{\small\ttfamily}}}
\makeatother
\def\bcj{\begin{conjecture}}
\def\ecj{\end{conjecture}}
\def\bcr{\begin{corollary}}
\def\ecr{\end{corollary}}
\def\bd{\begin{definition}}
\def\ed{\end{definition}}
\def\bea{\begin{eqnarray}}
\def\eea{\end{eqnarray}}
\def\bem{\begin{enumerate}}
\def\eem{\end{enumerate}}
\def\bex{\begin{example}}
\def\eex{\end{example}}
\def\bim{\begin{itemize}}
\def\eim{\end{itemize}}
\def\bl{\begin{lemma}}
\def\el{\end{lemma}}
\def\bpf{\begin{proof}}
\def\epf{\end{proof}}
\def\bpp{\begin{proposition}}
\def\epp{\end{proposition}}
\def\bqu{\begin{question}}
\def\equ{\end{question}}
\def\br{\begin{remark}}
\def\er{\end{remark}}
\def\bt{\begin{theorem}}
\def\et{\end{theorem}}

\def\btb{\begin{tabular}}
\def\etb{\end{tabular}}

\newcommand{\nc}{\newcommand}

 \nc{\bA}{{\bf A}} \nc{\bB}{{\bf B}} \nc{\bC}{{\bf C}}
 \nc{\bD}{{\bf D}} \nc{\bE}{{\bf E}} \nc{\bF}{{\bf F}}
 \nc{\bG}{{\bf G}} \nc{\bH}{{\bf H}} \nc{\bI}{{\bf I}}
 \nc{\bJ}{{\bf J}} \nc{\bK}{{\bf K}} \nc{\bL}{{\bf L}}
 \nc{\bM}{{\bf M}} \nc{\bN}{{\bf N}} \nc{\bO}{{\bf O}}
 \nc{\bP}{{\bf P}} \nc{\bQ}{{\bf Q}} \nc{\bR}{{\bf R}}
 \nc{\bS}{{\bf S}} \nc{\bT}{{\bf T}} \nc{\bU}{{\bf U}}
 \nc{\bV}{{\bf V}} \nc{\bW}{{\bf W}} \nc{\bX}{{\bf X}}
 \nc{\bZ}{{\bf Z}}

\nc{\cA}{{\cal A}} \nc{\cB}{{\cal B}} \nc{\cC}{{\cal C}}
\nc{\cD}{{\cal D}} \nc{\cE}{{\cal E}} \nc{\cF}{{\cal F}}
\nc{\cG}{{\cal G}} \nc{\cH}{{\cal H}} \nc{\cI}{{\cal I}}
\nc{\cJ}{{\cal J}} \nc{\cK}{{\cal K}} \nc{\cL}{{\cal L}}
\nc{\cM}{{\cal M}} \nc{\cN}{{\cal N}} \nc{\cO}{{\cal O}}
\nc{\cP}{{\cal P}} \nc{\cQ}{{\cal Q}} \nc{\cR}{{\cal R}}
\nc{\cS}{{\cal S}} \nc{\cT}{{\cal T}} \nc{\cU}{{\cal U}}
\nc{\cV}{{\cal V}} \nc{\cW}{{\cal W}} \nc{\cX}{{\cal X}}
\nc{\cZ}{{\cal Z}}

\nc{\hA}{{\hat{A}}} \nc{\hB}{{\hat{B}}} \nc{\hC}{{\hat{C}}}
\nc{\hD}{{\hat{D}}} \nc{\hE}{{\hat{E}}} \nc{\hF}{{\hat{F}}}
\nc{\hG}{{\hat{G}}} \nc{\hH}{{\hat{H}}} \nc{\hI}{{\hat{I}}}
\nc{\hJ}{{\hat{J}}} \nc{\hK}{{\hat{K}}} \nc{\hL}{{\hat{L}}}
\nc{\hM}{{\hat{M}}} \nc{\hN}{{\hat{N}}} \nc{\hO}{{\hat{O}}}
\nc{\hP}{{\hat{P}}} \nc{\hR}{{\hat{R}}} \nc{\hS}{{\hat{S}}}
\nc{\hT}{{\hat{T}}} \nc{\hU}{{\hat{U}}} \nc{\hV}{{\hat{V}}}
\nc{\hW}{{\hat{W}}} \nc{\hX}{{\hat{X}}} \nc{\hZ}{{\hat{Z}}}

\def\Dbar{\leavevmode\lower.6ex\hbox to 0pt
{\hskip-.23ex\accent"16\hss}D}
\begin{document}

\def\be{\begin{eqnarray}}
\def\ee{\end{eqnarray}}

\newcommand{\sa}{\mathscr{A}}
\newcommand{\sm}{\mathscr{M}}

\newcommand{\cfg}{\dot \fg}
\newcommand{\cFg}{\dot \Fg}
\newcommand{\ccg}{\dot \cg}
\newcommand{\circj}{\dot {\mathbf J}}
\newcommand{\circs}{\circledS}
\newcommand{\jmot}{\mathbf J^{-1}}

\newcommand{\rmd}{\mathrm d}
\newcommand{\mca}{\ ^-\ca}
\newcommand{\pca}{\ ^+\ca}
\newcommand{\peq}{^\Psi\!=}
\newcommand{\lt}{\left}
\newcommand{\rt}{\right}
\newcommand{\HN}{\hat{H}(N)}
\newcommand{\HM}{\hat{H}(M)}
\newcommand{\Hv}{\hat{H}_v}
\newcommand{\cyl}{\mathbf{Cyl}}
\newcommand{\lag}{\left\langle}
\newcommand{\rag}{\right\rangle}
\newcommand{\Ad}{\mathrm{Ad}}
\newcommand{\trace}{\mathrm{tr}}
\newcommand{\bbc}{\mathbb{C}}
\newcommand{\AC}{\overline{\mathcal{A}}^{\mathbb{C}}}
\newcommand{\Ar}{\mathbf{Ar}}
\newcommand{\uc}{\mathrm{U(1)}^3}
\newcommand{\M}{\hat{\mathbf{M}}}
\newcommand{\spin}{\text{Spin(4)}}
\newcommand{\id}{\mathrm{id}}
\newcommand{\Pol}{\mathrm{Pol}}
\newcommand{\Fun}{\mathrm{Fun}}
\newcommand{\bp}{p}
\newcommand{\act}{\rhd}
\newcommand{\data}{\lt(j_{ab},A,\bar{A},\xi_{ab},z_{ab}\rt)}
\newcommand{\datao}{\lt(j^{(0)}_{ab},A^{(0)},\bar{A}^{(0)},\xi_{ab}^{(0)},z_{ab}^{(0)}\rt)}
\newcommand{\deltadata}{\lt(j'_{ab}, A',\bar{A}',\xi_{ab}',z_{ab}'\rt)}
\newcommand{\background}{\lt(j_{ab}^{(0)},g_a^{(0)},\xi_{ab}^{(0)},z_{ab}^{(0)}\rt)}
\newcommand{\sgn}{\mathrm{sgn}}
\newcommand{\vth}{\vartheta}
\newcommand{\rmi}{\mathrm{i}}
\newcommand{\bfmu}{\pmb{\mu}}
\newcommand{\bfnu}{\pmb{\nu}}
\newcommand{\bfm}{\mathbf{m}}
\newcommand{\bfn}{\mathbf{n}}
\newcommand{\perk}{\mathfrak{S}_k}
\newcommand{\dens}{\mathrm{D}}
\newcommand{\iden}{\mathbb{I}}
\newcommand{\End}{\mathrm{End}}

\newcommand{\sz}{\mathscr{Z}}
\newcommand{\sk}{\mathscr{K}}

\title{Impact of Oxygen Vacancies in Josephson Junction on Decoherence of Superconducting Qubits}

\author{Hanqin Bai}
\affiliation{College of Physics and Electronic Engineering, Center for Computational Sciences,  Sichuan Normal University, Chengdu 610068, China}

\author{Shi-Yao Hou}
\email[]{hshiyao@sicnu.edu.cn}
\affiliation{College of Physics and Electronic Engineering, Center for Computational Sciences,  Sichuan Normal University, Chengdu 610068, China}

\author{Mu Lan}
\affiliation{Sichuan Province Key Laboratory of Optoelectronic Sensor Devices and Systems, College of Optoelectronic Engineering, Chengdu University of Information Technology, Chengdu 610225, China 
,Sichuan Meteorological Optoelectronic Sensor Technology and Application Engineering Research Center, Chengdu University of Information Technology, Chengdu 610225, China}

\date{\today}

\begin{abstract}

Superconducting quantum circuits are promising platforms for scalable quantum computing, where qubit coherence is critically determined by microscopic defects in the oxide tunneling barrier of Josephson junctions. Amorphous Al$_2$O$_3$ is widely used as a barrier material, but under irradiation, oxygen vacancy (V$_O$) defects are readily generated, introducing noise sources that accelerate qubit decoherence. We systematically investigate the structural characteristics and electronic impact of V$_O$ defects in amorphous Al$_2$O$_3$ using first-principles calculations and \textit{ab initio} molecular dynamics. Our results show that both the coordination environment and concentration of V$_O$s strongly influence electrical conductivity. In particular, two- and three-coordinated V$_O$s, unique to the amorphous structure, enhance conductivity more than conventional four-coordinated vacancies. Increasing V$_O$ concentration amplifies conductivity fluctuations, which we link to critical current noise in Josephson junctions. Using a noise model, we estimate that higher V$_O$ densities lead to shorter qubit coherence times. These findings provide insights for radiation-hard design of superconducting quantum devices.

\end{abstract}

\maketitle

\section{Introduction}

High-performance superconducting qubits are essential for realizing large-scale practical quantum computing~\cite{nielsen2010quantum}. In recent years, continuous improvements in gate fidelity and coherence times have enabled superconducting quantum processors to integrate hundreds of qubits, laying the foundation for scalable universal quantum computation~\cite{sanders2025superconducting}. Nevertheless, decoherence remains a key factor limiting long-term scalability. Extensive studies show that decoherence predominantly originates from defect-related states in materials, such as two-level systems (TLSs), charge noise, flux noise, radio-frequency energy leakage, and parasitic couplings~\cite{dielectricloss16,rf-squid17,quasiparticle18,quasiparticles-chare19,tls-lifetime-and-coherence20,fulx-noise21,tls-relaxtion22}. In particular, microscopic defects in device materials---especially in the Al$_2$O$_3$ insulating layer---introduce low-frequency noise that degrades qubit energy-level stability and gate fidelity. Impurity states associated with hydrogen-related defects~\cite{holder2013bulk25,O-H...Hhydrogen} and oxygen vacancies (V$_O$)~\cite{oxygen37,guo2016oxygen38,qiu2024manipulation39} have been identified as major contributors to such noise sources.

To mitigate decoherence, various strategies have been proposed, including suppressing the coupling between defects and the environment~\cite{envrionment-entagle}, as well as optimizing fabrication processes---for example, employing thermal annealing to reduce V$_O$ formation~\cite{annealing67,anneling} and thereby suppress noise. Previous studies demonstrate that hydrogen impurities can form O--H$\cdots$O tunneling configurations, giving rise to noise in dielectric materials~\cite{O-H...Hhydrogen}. In addition, hydrogen binds with aluminum vacancies to form V$_{\text{Al}}$--H complexes, which constitute dominant defects in oxide layers and significantly deteriorate material performance~\cite{holder2013bulk25}. These findings highlight the importance of understanding microscopic noise origins and defect structures for controlling decoherence in superconducting qubits.

Oxygen vacancies, in particular, play a crucial role in determining the electrical properties of oxide materials. Previous work shows that the microscopic structure of V$_O$s strongly influences charge transport behavior~\cite{qiu2024manipulation39,muller2019towards78}. Due to its excellent insulating properties, amorphous Al$_2$O$_3$ is commonly used as the oxide barrier in Josephson junctions. However, under high-energy irradiation, Al--O bonds in Al$_2$O$_3$ can break, leading to Al dangling bonds at metal-oxide interfaces and V$_O$s within the oxide bulk~\cite{ding2021damage49,ding2014total50,zhu2018total51}. X-ray photoelectron spectroscopy measurements indicate that V$_O$ concentration increases with irradiation dose. In addition, oxides may trap excess charges, altering local charge density and consequently modifying material properties. It has also been reported that variations in the position, distribution, and concentration of V$_O$s within the tunnel barrier significantly affect the local density of states (LDOS), while Coulomb repulsion between Al ions near vacancies promotes electron delocalization and modifies electrical conductivity. These studies reveal that V$_O$s can alter electron tunneling pathways and conductive channels in Josephson junctions, thereby influencing transport behavior.

Despite these advances, direct investigations of the electronic structure of irradiation-induced V$_O$s in amorphous oxides, as well as their impact on transport properties and qubit decoherence mechanisms, remain limited. A systematic understanding of how irradiation-induced V$_O$s in amorphous Al$_2$O$_3$ affect charge transport and coherence is therefore highly desirable for improving coherence and radiation tolerance of Al/AlO$_X$/Al Josephson junctions.

In this work, we employ density functional theory (DFT) to investigate the electronic structure of irradiation-induced V$_O$s in amorphous Al$_2$O$_3$ and explain how the coordination environment and concentration of vacancies influence charge transport properties. We first construct realistic amorphous models via \textit{ab initio} molecular dynamics and validate their structural properties. We then systematically analyze the effect of V$_O$ coordination (2-, 3-, and 4-fold) and concentration on electronic structure and conductivity. Finally, we link the conductivity fluctuations to critical current noise in Josephson junctions and provide quantitative estimates of qubit decoherence times. Our results demonstrate that both coordination environment and concentration of V$_O$s are critical factors determining qubit coherence, offering guidance for material design and defect engineering in superconducting quantum devices.

\section{Theoretical Details}
\label{sec:td}

To construct an amorphous Al$_2$O$_3$ structure, we employed a melt-quench procedure via \textit{ab initio} molecular dynamics (AIMD) simulations using the Vienna ab initio simulation package (VASP). This approach effectively reproduces, at the atomic scale, the experimental process of obtaining amorphous materials through high-temperature melting followed by rapid quenching, thereby providing a reliable model for investigating their local structural and electronic properties. Using a supercell containing 64 Al atoms and 96 O atoms, all simulations were performed in the NVT ensemble with the Nose-Hoover thermostat, using an integration time step of 0.5 fs. First, the system temperature was rapidly increased from 300 K to 4000 K in 100 K increments at a heating rate of 100 K/ps to melt the crystalline structure. Subsequently, the temperature was maintained at 4000 K for 0.5 ps. Finally, the system was cooled from 4000 K to 300 K at a cooling rate of 100 K/ps to obtain the amorphous Al$_2$O$_3$ structure. 

Spin-polarized first-principles calculations were performed within the density functional theory (DFT) framework using VASP~\cite{VASP_kresse1993ab,Vasp_kresse1996efficient} and projector augmented wave (PAW)~\cite{PAW_blochl1994projector,PAW_kresse1999ultrasoft} pseudopotentials. The exchange-correlation interactions were treated with the strongly constrained and appropriately normed semilocal density functional (SCAN)~\cite{Scan_sun2015strongly} meta-generalized gradient approximation (metaGGA) for geometry optimization, as it provides an accurate description of bond lengths and coordination in amorphous oxides. To address the underestimation of bandgaps, electronic band structures and densities of states were calculated using the HSE06~\cite{Hse_heyd2003hybrid,Hse_ge2006erratum} hybrid functional, which incorporates 45\% nonlocal Hartree-Fock exchange. The electronic configurations were set as 3s$^2$3p$^1$ for Al and 2s$^2$2p$^4$ for O. A kinetic energy cutoff of 520 eV was employed. Electronic energy minimization was achieved with a tolerance of $10^{-6}$ eV, while ionic relaxation proceeded until the force on each atom was below $0.02~\text{eV}/\mathring{\text{A}} $.

\section{Model Construction and Structural Validation}
\label{sec:struct}

The constructed amorphous Al$_2$O$_3$ supercell contains 160 atoms (64 Al and 96 O), with a total volume of 1341 $\mathring{\text{A}}^3$ and a density of 4.04 g/cm$^3$. The calculated band gap of the amorphous Al$_2$O$_3$ model is approximately 6.84 eV, which is consistent with experimental band gaps reported for amorphous Al$_2$O$_3$ films (6--7 eV)~\cite{band_gap_7-0.1,amphous_band_gap_6-7}. To investigate the structural characteristics and verify the reliability of the model, we analyzed the pair correlation function (PCF), bond lengths, and coordination distribution. The final ten configurations from the molecular dynamics trajectories were averaged, and the averaged atomic positions were used to compute the atomic spatial correlations.

The pair correlation functions $g_{ab}(r)$ in a binary system are defined such that, sitting on one atom of species $a$, the probability of finding one atom of species $b$ in a spherical shell between $r$ and $r+\Delta r$ is
\[
\langle n_{a,b}(r, r+\Delta r) \rangle = \rho_b \, 4 \pi r^2 \, g_{a,b}(r) \, \Delta r,
\]
where $\rho_b = N_b/V$ is the number density of species $b$, and $N_b$ is the total number of atoms of species $b$~\cite{gutierrez2002molecular}.

\begin{figure*}[htbp]
\centering
\begin{subfigure}{0.45\textwidth}
\centering
\includegraphics[width=\linewidth]{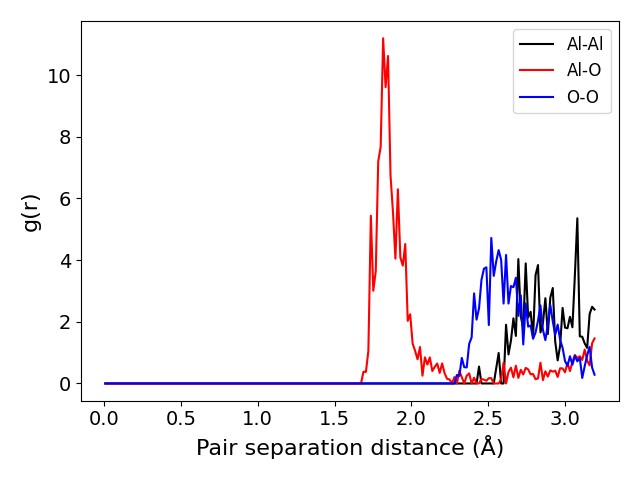}
\caption{Total pair correlation function for amorphous Al$_2$O$_3$.}
\label{fig:31a}
\end{subfigure}
\hfill
\begin{subfigure}{0.45\textwidth}
\centering
\includegraphics[width=\linewidth]{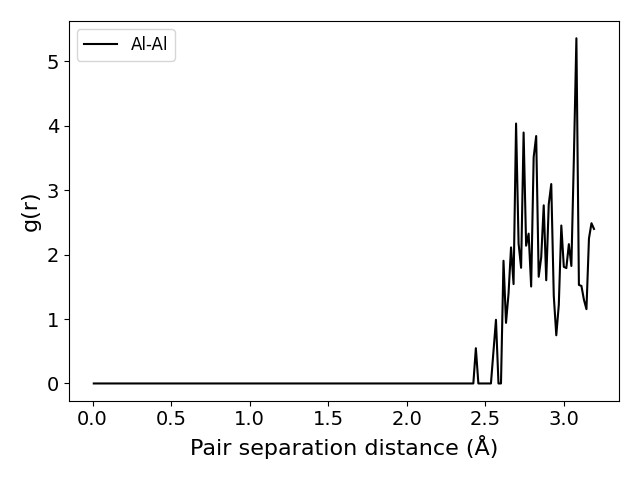}
\caption{Al--Al pair correlation function.}
\label{fig:31b}
\end{subfigure}
\vfill
\begin{subfigure}{0.45\textwidth}
\centering
\includegraphics[width=\linewidth]{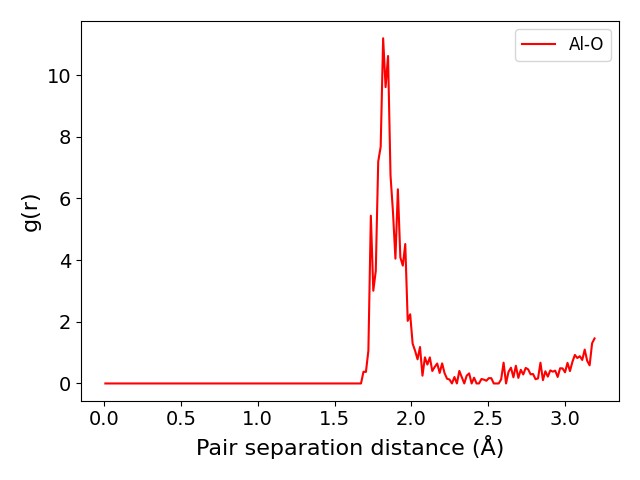}
\caption{Al--O pair correlation function.}
\label{fig:31c}
\end{subfigure}
\hfill
\begin{subfigure}{0.45\textwidth}
\centering
\includegraphics[width=\linewidth]{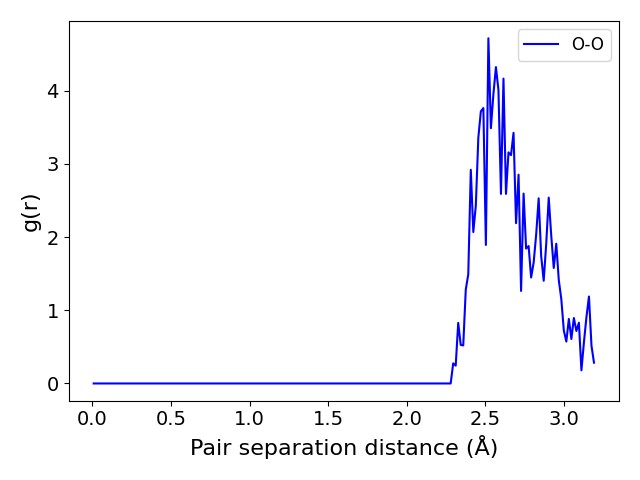}
\caption{O--O pair correlation function.}
\label{fig:31d}
\end{subfigure}
\caption{Partial pair correlation functions for amorphous Al$_2$O$_3$.}
\end{figure*}

In Fig.~\ref{fig:31a}, the PCF of amorphous Al$_2$O$_3$ exhibits a pronounced first peak, corresponding to the nearest-neighbor atomic correlations. This indicates distinct short-range order in the system, primarily arising from the strong Al--O bonding within the network structure. Beyond the first peak, the intensity gradually decreases and becomes significantly broadened, reflecting the long-range structural disorder characteristic of the amorphous phase.
Figs.~\ref{fig:31b}--\ref{fig:31d} display the Al--Al, Al--O, and O--O pair correlation functions, respectively. Based on the positions of the first peaks, the estimated nearest-neighbor distances are approximately $3.08 \pm 0.05~\mathring{\text{A}}$ for Al--Al, $1.81 \pm 0.10~\mathring{\text{A}}$ for Al--O, and $2.52 \pm 0.20~\mathring{\text{A}}$ for O--O, with uncertainties estimated from the half-width at half-maximum (HWHM).

A useful supplementary information can be obtained by integrating around the first peak in the PCF, which provides the average coordination number $n_{ab}$~\cite{structure}:
\[
n_{ab}(R) = 4 \pi \rho_b \int_0^R g_{ab}(r) \, r^2 \, dr,
\]
where $R$ is a cutoff, usually chosen as the position of the minimum after the first peak of $g_{ab}(r)$. Within the cutoff distances $R_{\text{Al--Al}} = 3.3~\mathring{\text{A}}$, $R_{\text{Al--O}} = 2.6~\mathring{\text{A}}$, and $R_{\text{O--O}} = 3.2~\mathring{\text{A}}$, the results show that each Al atom has, on average, 6.62 Al neighbors and 5.78 O neighbors, while each O atom has an average of 3.85 Al neighbors and 11.08 O neighbors. 

\begin{figure*}[htbp]
\centering
\begin{subfigure}{0.45\textwidth}
\centering
\includegraphics[width=\linewidth]{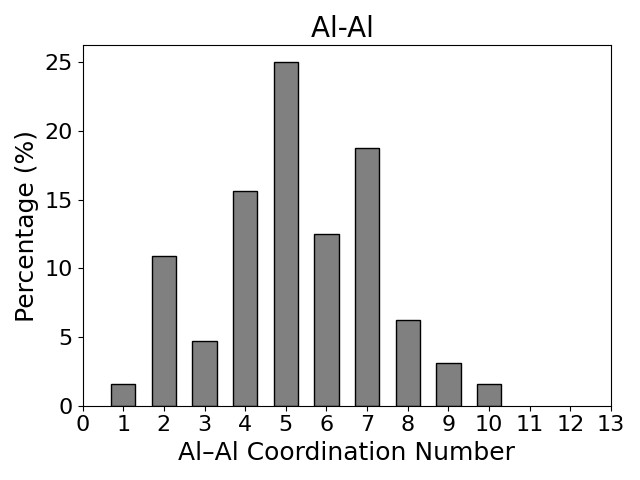}
\caption{Distribution of Al coordination numbers (Al--O).}
\label{fig:32a}
\end{subfigure}
\hfill
\begin{subfigure}{0.45\textwidth}
\centering
\includegraphics[width=\linewidth]{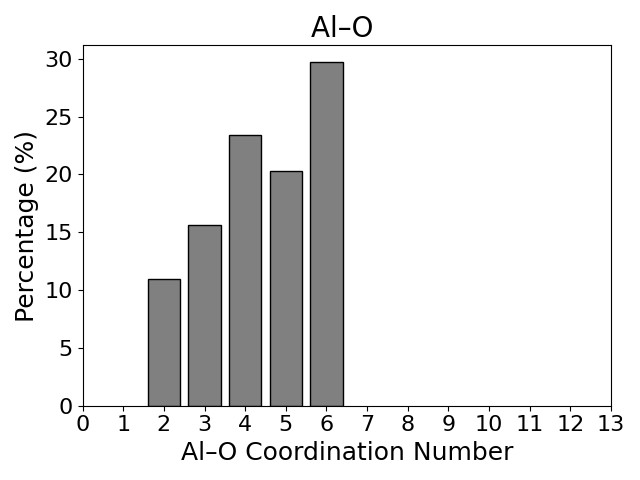}
\caption{Distribution of O coordination numbers (O--Al).}
\label{fig:32b}
\end{subfigure}
\vfill
\begin{subfigure}{0.45\textwidth}
\centering
\includegraphics[width=\linewidth]{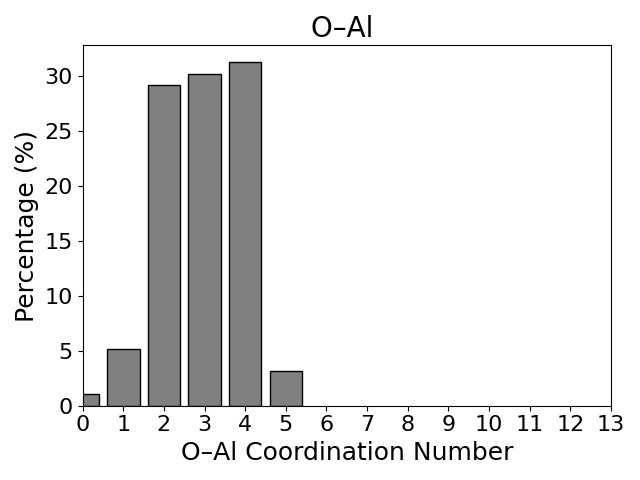}
\caption{Distribution of Al--Al coordination numbers.}
\label{fig:32c}
\end{subfigure}
\hfill
\begin{subfigure}{0.45\textwidth}
\centering
\includegraphics[width=\linewidth]{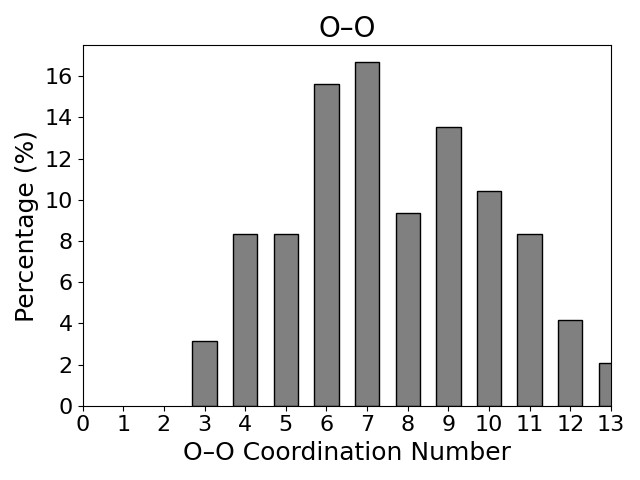}
\caption{Distribution of O--O coordination numbers.}
\label{fig:32d}
\end{subfigure}
\caption{Distribution of Al and O nearest-neighbor coordination in amorphous Al$_2$O$_3$.}
\end{figure*}

We further analyzed the coordination number distributions for different atoms and plotted the corresponding histograms (Figs.~\ref{fig:32a}--\ref{fig:32d}). For Al atoms, those with coordination numbers of 4, 5, and 6 account for approximately 74\% of the total, with sixfold-coordinated Al atoms being the most abundant, indicating that octahedral configurations are dominant. Al atoms with coordination numbers of 2 or 3 constitute only about 15\%. For O atoms, the coordination number distribution shows that 29\% are twofold-coordinated, 30\% are threefold-coordinated, and 31\% are fourfold-coordinated, while only about 10\% of O atoms are onefold or fivefold coordinated. The coordination number distributions of Al--Al and O--O pairs are relatively broad, with coordination numbers of 5 and 7 being the most frequent for Al--Al, and 6, 7, 9, and 10 for O--O. The observed deviations may arise from the accuracy of the cutoff distance.

Lamparter's experimental measurements~\cite{lamparter1997structure} of amorphous Al$_2$O$_3$ reported Al--O, Al--Al, and O--O bond lengths of 1.8, 3.2, and 2.8 $\mathring{\text{A}}$, respectively, with an Al coordination number of 4.1. Oka's study~\cite{oka1979structural} on anodically oxidized amorphous Al$_2$O$_3$ revealed that, in low-density porous films, the Al--O bond length is about 1.8 $\mathring{\text{A}}$ and Al atoms are tetrahedrally coordinated, while in high-density nonporous films, the Al--O bond length increases to about 1.9 $\mathring{\text{A}}$ and Al atoms become octahedrally coordinated.

Table~\ref{Tab:1} summarizes and compares our simulation results with available experimental data. Experiments generally show Al--O bond lengths ranging from 1.8 to 1.9 $\mathring{\text{A}}$ and Al coordination numbers between 4.1 and 4.8. It is widely accepted that the amorphous Al$_2$O$_3$ network consists primarily of tetrahedral (AlO$_4$) and octahedral (AlO$_6$) polyhedral units, with their relative proportions depending on the preparation method. Experimental evidence indicates that both coordination number and polyhedral proportion are correlated with sample density. Gonzalo and Börje~\cite{gutierrez2002molecular} conducted simulations over a density range of 3.0--3.3 g/cm$^3$ and found that, although bond lengths and bond angle distributions show minor variations, the coordination and ring statistics change markedly with density: as density increases, the number of overcoordinated polyhedra (AlO$_5$ and AlO$_6$) rises, whereas undercoordinated polyhedra (AlO$_3$ and AlO$_4$) decrease. Therefore, our results are consistent with experimental and theoretical observations~\cite{structure,amphous_band_gap_6-7}.

\renewcommand{\arraystretch}{1.5} 
\begin{table}[t]
\centering
\caption{Comparison of our simulation results with available experimental data for amorphous Al$_2$O$_3$. Bond lengths $r$ (in $\mathring{\text{A}}$) and coordination numbers $n$ are given for each atomic pair.}
\resizebox{\linewidth}{!}{%
\begin{tabular}{lcccccccc}
\hline
 & \multicolumn{2}{c}{Lamparter and Kniep} & \multicolumn{2}{c}{Oka} & \multicolumn{2}{c}{Gonzalo and Börje} & \multicolumn{2}{c}{This work} \\
\cline{2-9}
 & $r$ ($\mathring{\text{A}}$) & $n$ & $r$ ($\mathring{\text{A}}$) & $n$ & $r$ ($\mathring{\text{A}}$) & $n$ & $r$ ($\mathring{\text{A}}$) & $n$ \\
\hline
Al--O & $1.8 \pm 0.21$ & 4.1 & 1.85 & 4.64/4.81 & $1.76 \pm 0.1$ & 4.25 & $1.81 \pm 0.1$ & 5.78 \\
O--O  & $2.8 \pm 0.58$ & 8.5 & -- & -- & $2.75 \pm 0.2$ & 9.47 & $2.52 \pm 0.2$ & 11.08 \\
Al--Al& $3.2 \pm 0.55$ & 6   & -- & -- & $3.1 \pm 0.21$ & 8.26 & $3.08 \pm 0.05$ & 6.62 \\
\hline
\end{tabular}%
}
\label{Tab:1}
\end{table}

\section{Electronic Structure and Transport Properties}
\label{sec:results}

Based on H.P. Zhu~\cite{zhu2018total51} and Man Ding's~\cite{ding2021damage49,ding2014total50} findings, oxygen vacancies are generated in Al$_2$O$_3$ under irradiation, with vacancy concentration increasing with dose~\cite{ding2021damage49}. Here, vacancies are assumed to be irradiation-induced rather than intrinsic or fabrication-related.

\subsection{Influence of V$_O$ Coordination Number on Electrical Conductivity}

Unlike crystalline Al$_2$O$_3$, the structure of amorphous Al$_2$O$_3$ features regional disorder, and irradiation often induces V$_O$s with different coordination numbers. The coordination number of a V$_O$ not only determines the depth of the associated defect levels but also affects the strength of the local electrostatic potential perturbation and its capability to modulate charge carrier transport. To systematically investigate how the coordination environment of V$_O$s affects electrical conductivity, we constructed specific V$_O$ defect models by removing oxygen atoms with different coordination numbers (2-fold, 3-fold, and 4-fold coordination) in amorphous Al$_2$O$_3$. To eliminate structural bias, oxygen vacancies of each type were generated through random selection while keeping all other structural conditions identical, enabling a direct comparison of the independent effects of vacancy coordination on electrical conductivity. The electronic structure was calculated using the Tran-Blaha modified Becke-Johnson (TB-MBJ) exchange potential. Based on the obtained electronic band structure, the conductivity was evaluated using the semiclassical Boltzmann transport theory within the constant relaxation time approximation. Note that within this approximation, the calculated conductivity $\sigma$ is proportional to the relaxation time $\tau$, and we report the quantity $\sigma/\tau$ (with units $\Omega^{-1} \mathrm{m}^{-1} \mathrm{s}^{-1}$), which reflects the intrinsic electronic transport properties apart from scattering time.

\begin{figure}[htbp]
\centering
\includegraphics[scale=0.27]{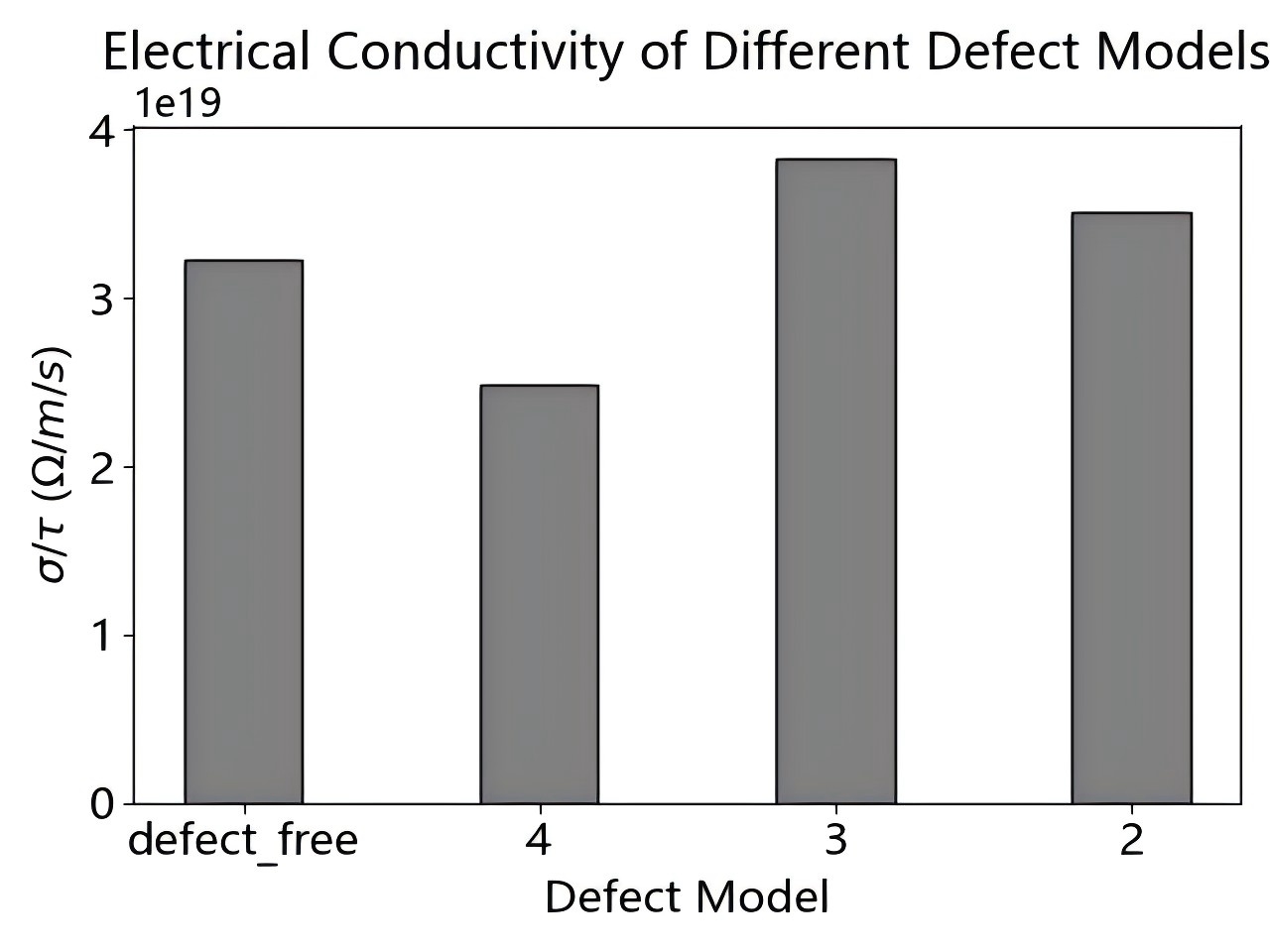}
\caption{Electrical conductivity ( $\sigma/\tau$) comparison of amorphous Al$_2$O$_3$ models with different V$_O$ coordination environments.}
\label{fig:411}
\end{figure}

In the three models, randomly selected oxygen atoms with coordination numbers of 4, 3, and 2 were removed, respectively. We then calculated the electrical conductivity of each model. As shown in Fig.~\ref{fig:411}, when the number of vacancies remains constant, the influence of different coordination numbers on conductivity is significant. The defect-free model has a conductivity of $3.23 \times 10^{19}~\Omega^{-1} \mathrm{m}^{-1} \mathrm{s}^{-1}$. After introducing a 3-coordinated V$_O$, the conductivity increases significantly to $3.83 \times 10^{19}~\Omega^{-1} \mathrm{m}^{-1} \mathrm{s}^{-1}$, a value comparable to that of the 2-coordinated vacancy model. Conversely, removing a 4-coordinated oxygen atom results in a significant reduction in conductivity to $2.48 \times 10^{19}~\Omega^{-1} \mathrm{m}^{-1} \mathrm{s}^{-1}$. This may be attributed to the fact that, in crystalline Al$_2$O$_3$, oxygen atoms predominantly adopt a 4-fold coordination, and most oxygen atoms in the constructed amorphous structure also maintain this coordination. As a result, introducing a vacancy at a 4-coordinated site tends to preserve the material's insulating nature.

\begin{figure*}[htbp]
\centering
\begin{subfigure}{0.45\textwidth}
\centering
\includegraphics[width=\linewidth]{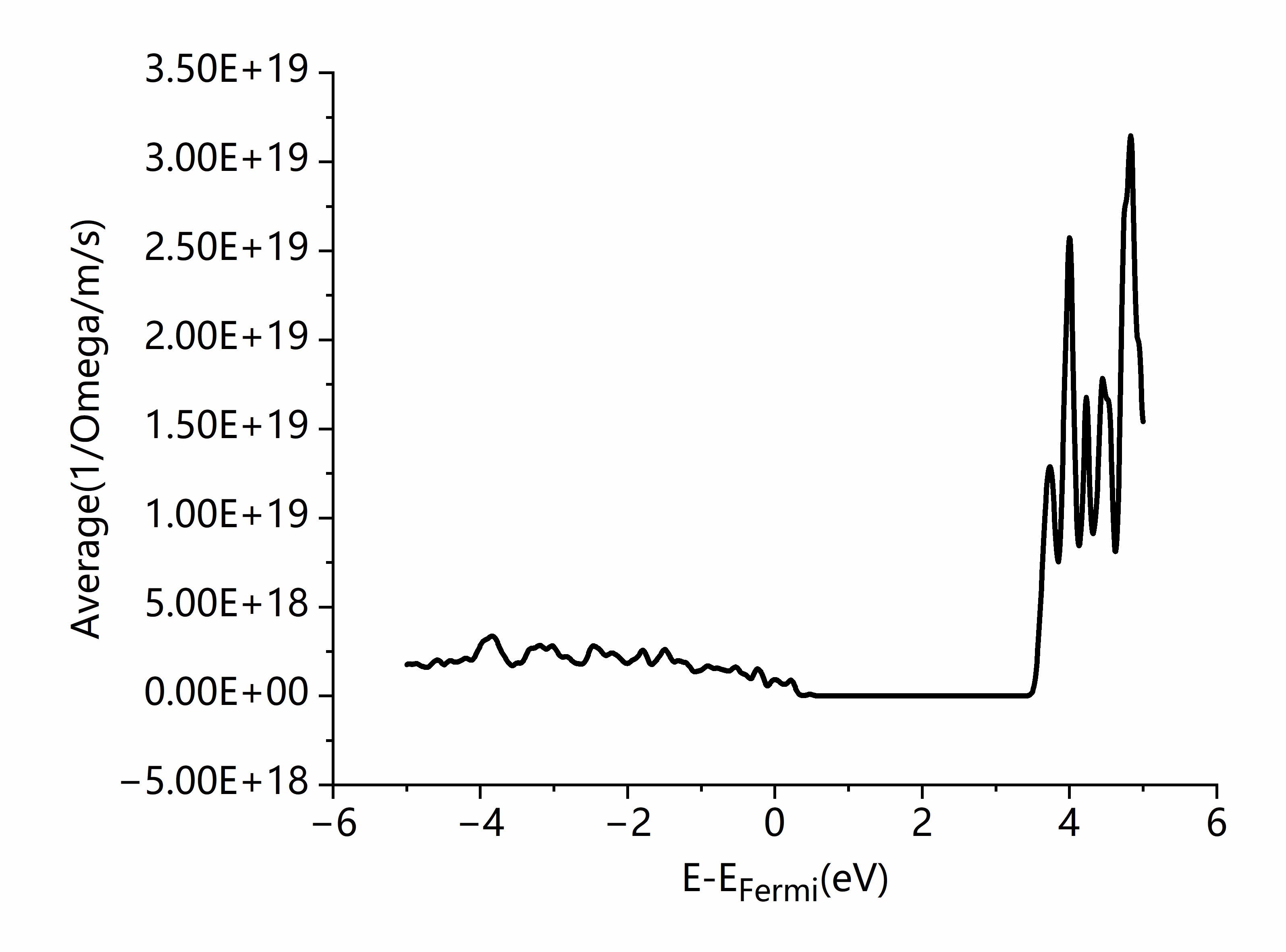}
\caption{Defect-free model.}
\label{fig:312a}
\end{subfigure}
\hfill
\begin{subfigure}{0.45\textwidth}
\centering
\includegraphics[width=\linewidth]{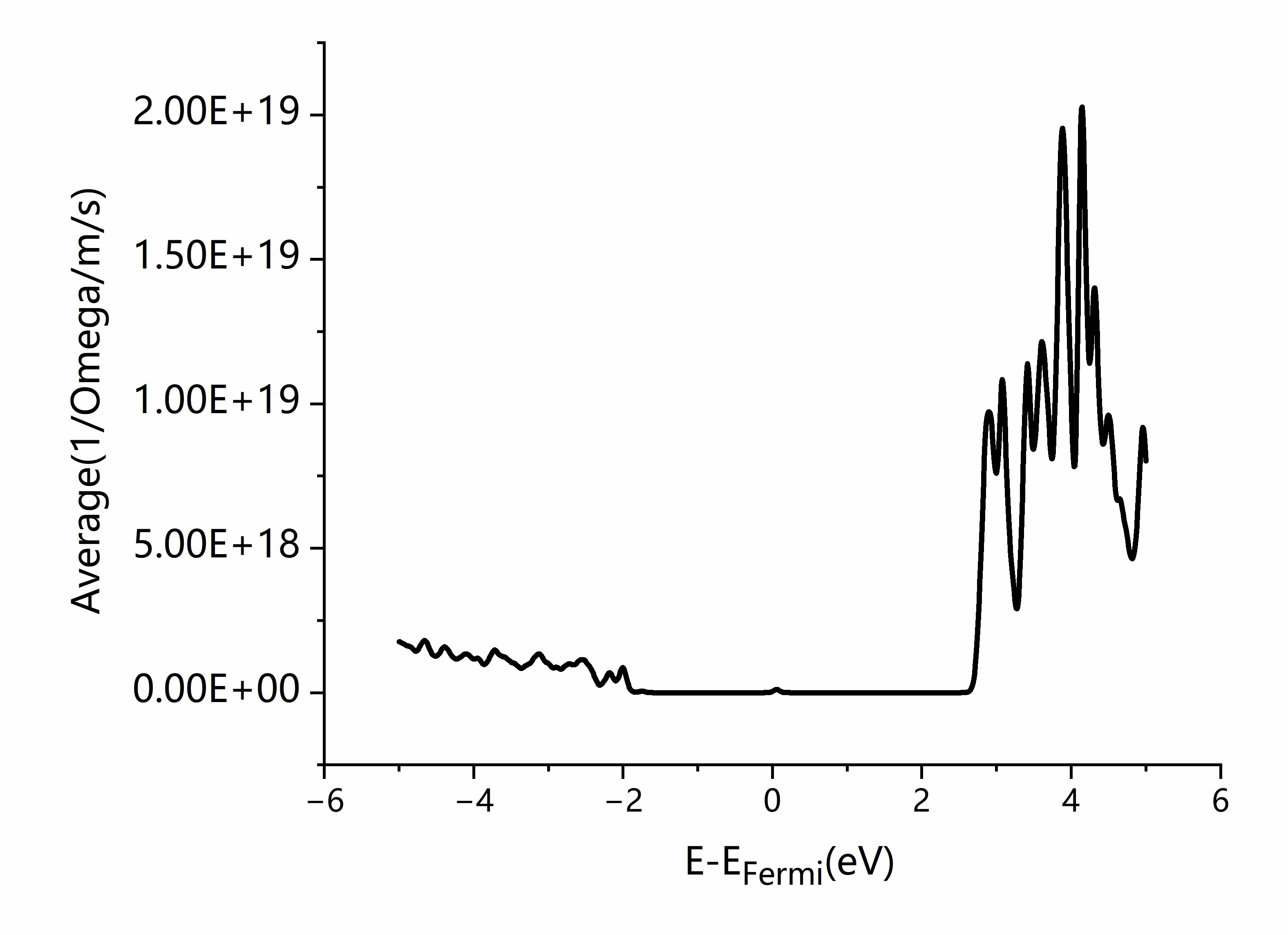}
\caption{4-coordinated V$_O$ model.}
\label{fig:312b}
\end{subfigure}
\vfill
\begin{subfigure}{0.45\textwidth}
\centering
\includegraphics[width=\linewidth]{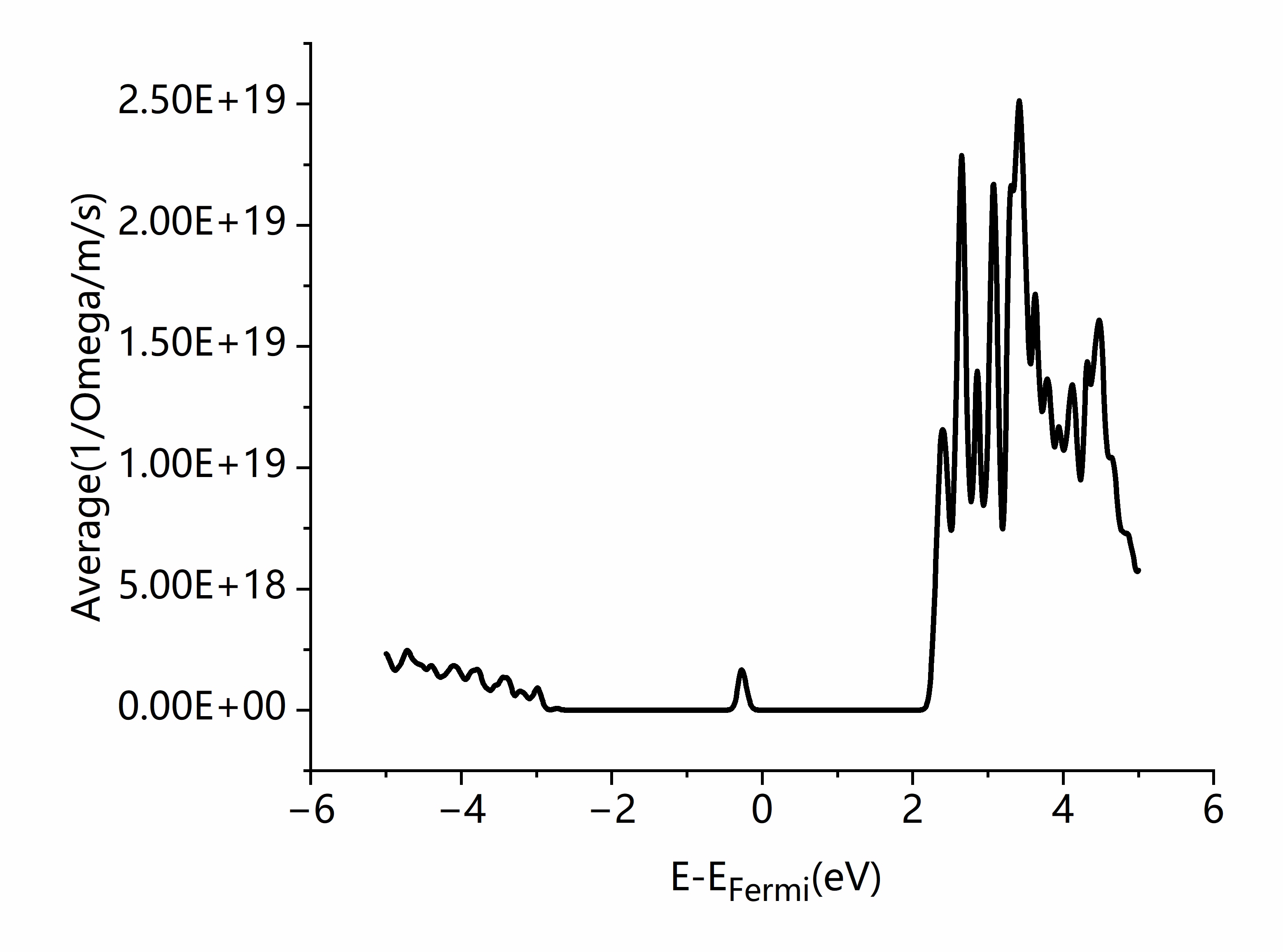}
\caption{3-coordinated V$_O$ model.}
\label{fig:312c}
\end{subfigure}
\hfill
\begin{subfigure}{0.45\textwidth}
\centering
\includegraphics[width=\linewidth]{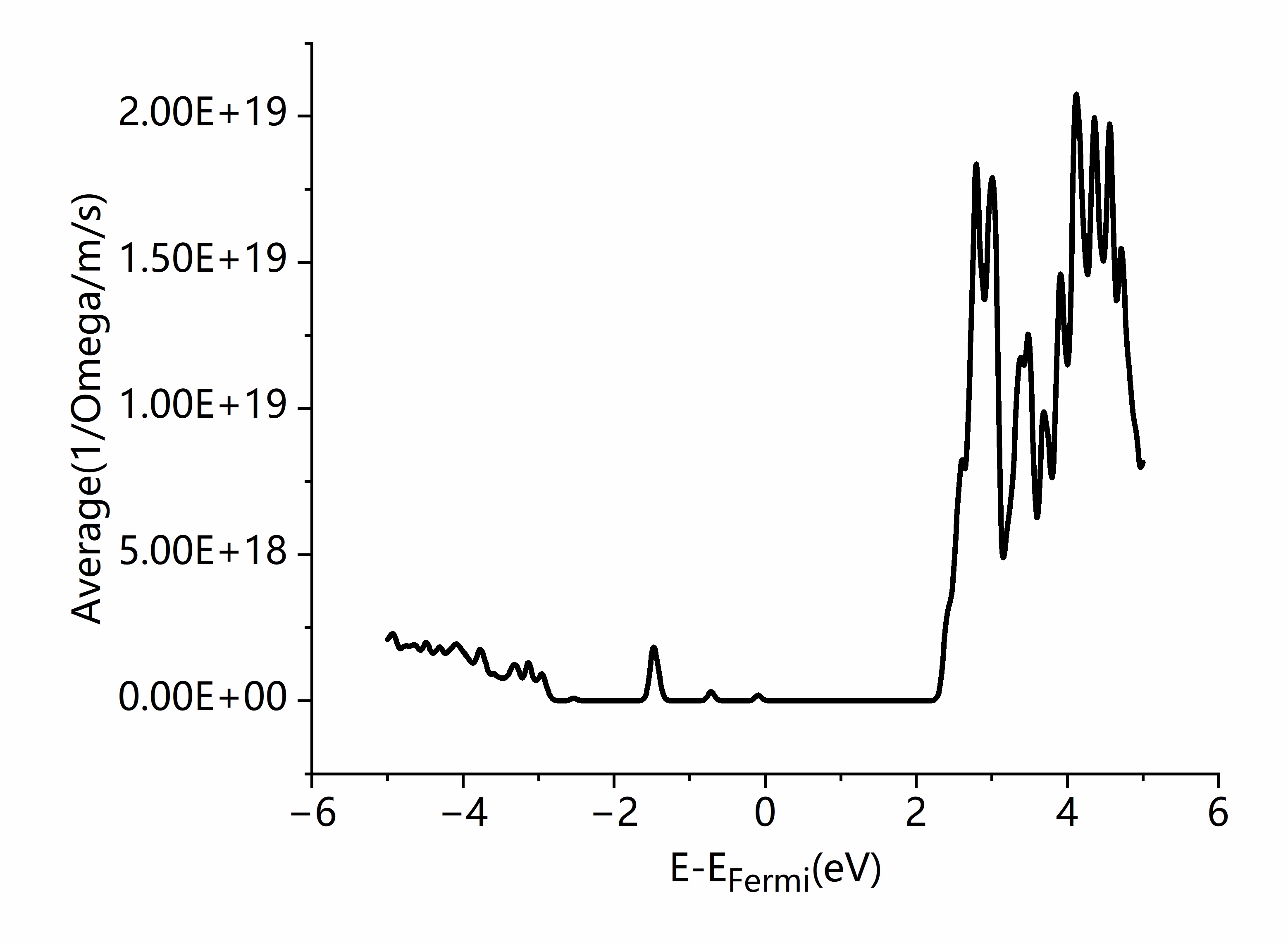}
\caption{2-coordinated V$_O$ model.}
\label{fig:312d}
\end{subfigure}
\caption{Electrical conductivity spectra ( $\sigma/\tau$) for amorphous Al$_2$O$_3$ models with different V$_O$ coordination environments.}
\label{fig:312}
\end{figure*}

We further compared the conductivity spectra of the four models (Fig.~\ref{fig:312}) and observed significant differences in the low-frequency region among the V$_O$ models with different coordination. Specifically, the conductivity spectrum of the 4-coordinated V$_O$ model remains relatively flat at low frequencies, with only a weak peak appearing near 0 eV (Fig.~\ref{fig:312b}). This indicates that the defect states introduced by 4-coordinated V$_O$s are relatively deep or highly localized, resulting in a negligible contribution to the overall electrical conductivity. In contrast, the 3-coordinated model exhibits a notable increase in conductivity in the low-frequency region ($<0.1$~eV) (Fig.~\ref{fig:312c}). Moreover, the 2-coordinated model shows multiple peaks in the low-frequency range (Fig.~\ref{fig:312d}), indicating that the defects may introduce shallow donor states, providing multiple effective channels for the generation of free charge carriers.

\begin{figure}[htbp]
\centering
\includegraphics[scale=0.65]{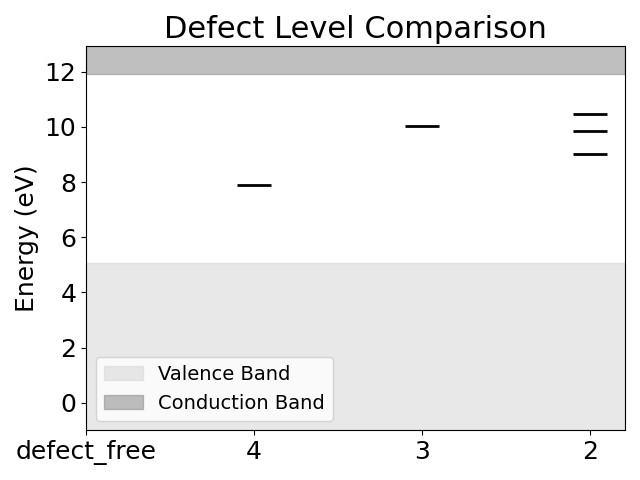}
\caption{Defect level distributions for amorphous Al$_2$O$_3$ models with different V$_O$ coordination environments.}
\label{fig:313}
\end{figure}

To clarify the physical origin of the observed conductivity trends, we further compare the defect level distributions of the four models, as shown in Fig.~\ref{fig:313}. In the 4-coordinated oxygen vacancy model, the defect level is located closer to the valence band, corresponding to a deep donor state with strong electron localization. Such localized states contribute negligibly to electrical conduction and may even act as hole traps, resulting in a conductivity lower than that of the defect-free model. In contrast, the chemical environments of 2- and 3-coordinated oxygen atoms, which are unique to the amorphous structure, are intrinsically less stable. Upon vacancy formation, these defects introduce shallow donor levels within the band gap. In particular, the defect level associated with the 3-coordinated vacancy lies closer to the conduction band minimum, while the 2-coordinated model introduces multiple defect levels, some of which are also located near the conduction band minimum. These features indicate a greater tendency to release free electrons and provide additional carrier transport pathways, thereby enhancing electrical conductivity. Nevertheless, compared with the 3-coordinated model, the 2-coordinated model also contains relatively deeper defect levels. The electronic structure results are consistent with the observed conductivity trends.

\begin{figure*}[htbp]
\centering
\begin{subfigure}{0.425\textwidth}
\includegraphics[width=\linewidth]{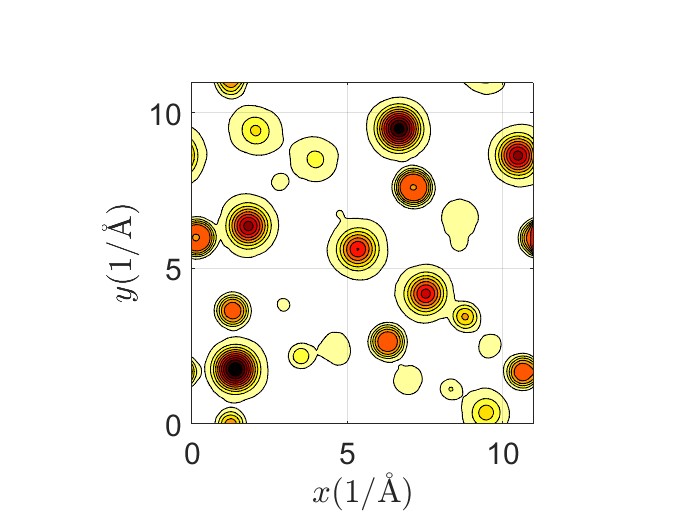}
\caption{4-coordinated O atom (defect-free).}
\label{fig:314a}
\end{subfigure}
\begin{subfigure}{0.425\textwidth}
\includegraphics[width=\linewidth]{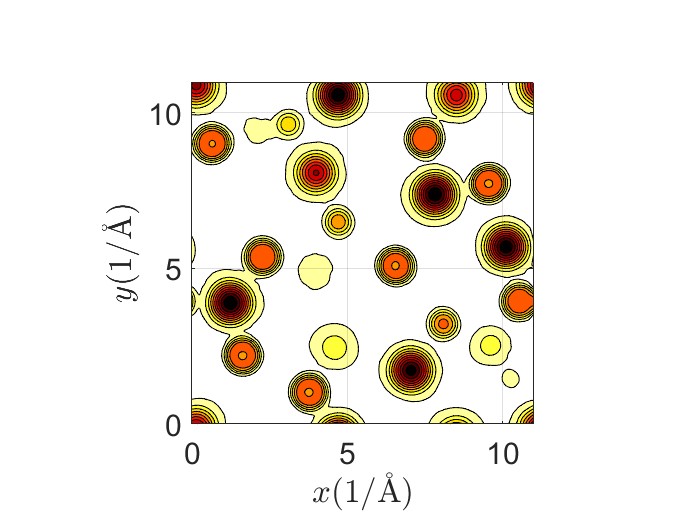}
\caption{After removing 4-coordinated O atom.}
\label{fig:314b}
\end{subfigure} \\
\begin{subfigure}{0.425\textwidth}
\includegraphics[width=\linewidth]{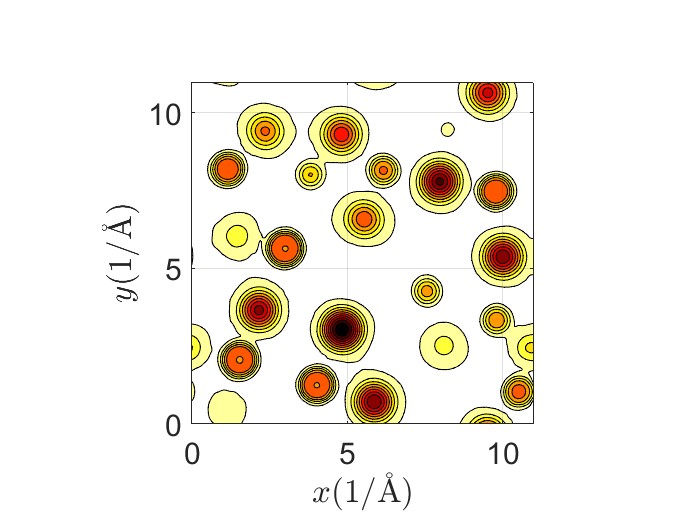}
\caption{3-coordinated O atom (defect-free).}
\label{fig:314c}
\end{subfigure}
\begin{subfigure}{0.425\textwidth}
\includegraphics[width=\linewidth]{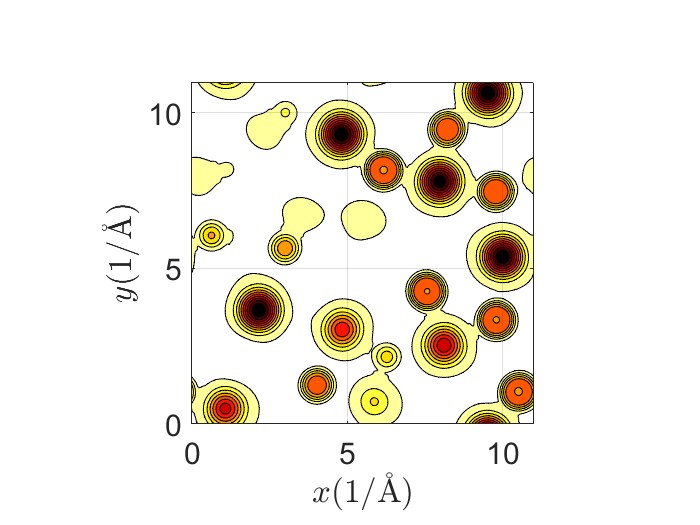}
\caption{After removing 3-coordinated O atom.}
\label{fig:314d}
\end{subfigure} \\
\begin{subfigure}{0.425\textwidth}
\includegraphics[width=\linewidth]{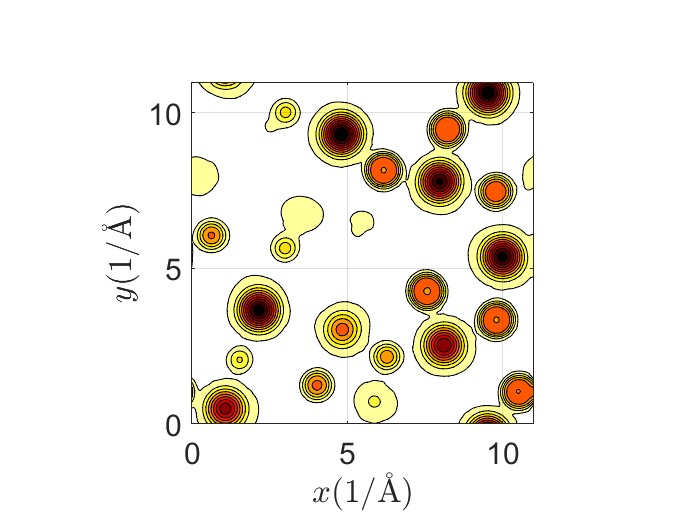}
\caption{2-coordinated O atom (defect-free).}
\label{fig:314e}
\end{subfigure}
\begin{subfigure}{0.425\textwidth}
\includegraphics[width=\linewidth]{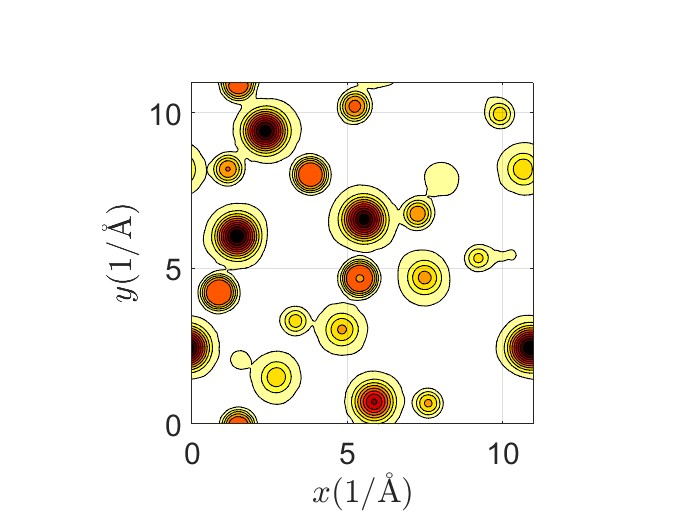}
\caption{After removing 2-coordinated O atom.}
\label{fig:314f}
\end{subfigure}
\caption{Electrostatic potential slices for different models ($x=0$ plane). (a), (c), (e) show electrostatic potential slice centered at the position of the oxygen atom before defect introduction. (b), (d), (f) show electrostatic potential slice near the defect site, centered on the atom nearest to the introduced vacancy.}
\label{fig:314}
\end{figure*}

To further investigate how oxygen vacancies regulate the local electrostatic potential distribution in amorphous Al$_2$O$_3$, we performed planar-slice visualizations of the potential near the defects. Fig.~\ref{fig:314a} shows the local electrostatic environment of a 4-coordinated oxygen atom in the defect-free model, while Fig.~\ref{fig:314b} shows the potential slice after removing this oxygen atom, using the nearest neighbor atom to the defect site as the center, with the $x = 0$ plane as the slicing plane. A comparison of these two panels reveals that, in the absence of defects, the equipotential contours are relatively symmetric and form regular circular potential wells. After the introduction of the V$_O$, the potential well near the defect remains approximately circular, while regions around Al atoms exhibit overlapping potential wells. Meanwhile, the potential depth at local Al and O sites increases, indicating a reconstruction of the local electrostatic potential induced by the defect. Figs.~\ref{fig:314c} and \ref{fig:314d} display the electrostatic potential slices of a 3-coordinated oxygen atom in the defect-free model and after vacancy formation, respectively (with $x = 0$ as the slicing plane). In the defect-free case, the equipotential contours are symmetric and circular. Upon the introduction of a 3-coordinated V$_O$, the equipotential contours evolve into irregular potential-well patterns, particularly around oxygen sites, where the equipotential lines become highly distorted and the depth of the local potential wells is reduced. In contrast, overlapping potential wells and an increased potential depth are observed around local Al atoms, which may originate from enhanced Coulomb repulsion between neighboring Al ions. This behavior reflects an overall enhancement of the local Coulomb potential and an increase in its spatial gradient.
For the 2-coordinated oxygen vacancy, shown in Figs.~\ref{fig:314e} and \ref{fig:314f}, the introduction of the defect leads to pronounced irregular distortions of the equipotential contours around oxygen atoms, accompanied by a reduction in the local potential well depth and an increased potential difference relative to surrounding regions. Such modifications of the electrostatic potential landscape can increase the probability of electron transitions when carriers traverse the defective barrier, thereby altering the transport properties of the system. 

\begin{figure*}[htbp]
\centering
\begin{subfigure}{0.225\textwidth}
\includegraphics[width=\linewidth]{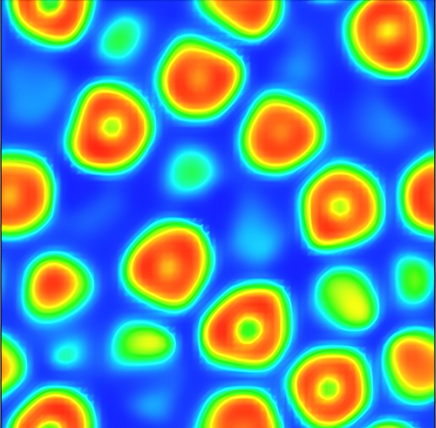}
\caption{4-coordinated O atom (defect-free).}
\label{fig:315a}
\end{subfigure}
\begin{subfigure}{0.225\textwidth}
\includegraphics[width=\linewidth]{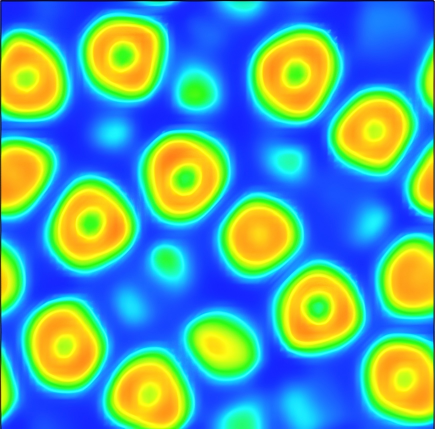}
\caption{After removing 4-coordinated O atom.}
\label{fig:315b}
\end{subfigure} \\
\begin{subfigure}{0.225\textwidth}
\includegraphics[width=\linewidth]{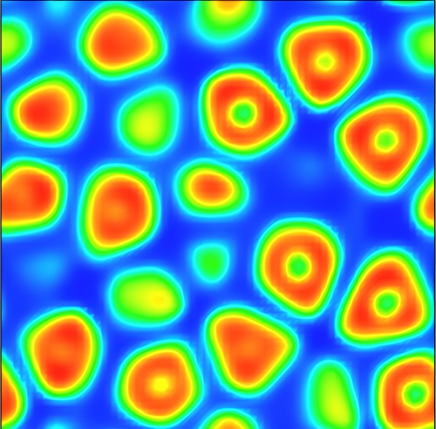}
\caption{3-coordinated O atom (defect-free).}
\label{fig:315c}
\end{subfigure}
\begin{subfigure}{0.225\textwidth}
\includegraphics[width=\linewidth]{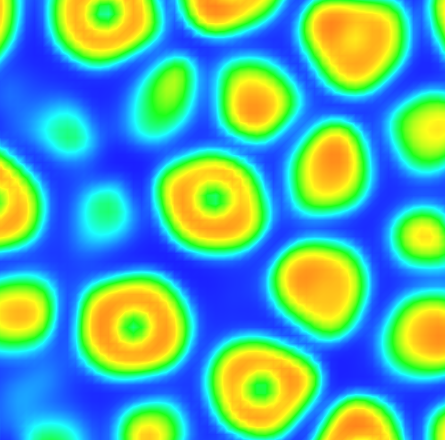}
\caption{After removing 3-coordinated O atom.}
\label{fig:315d}
\end{subfigure} \\
\begin{subfigure}{0.225\textwidth}
\includegraphics[width=\linewidth]{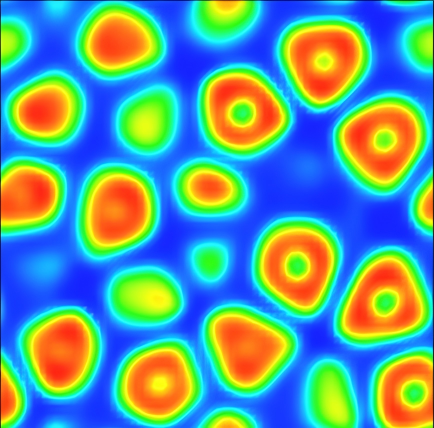}
\caption{2-coordinated O atom (defect-free).}
\label{fig:315e}
\end{subfigure}
\begin{subfigure}{0.225\textwidth}
\includegraphics[width=\linewidth]{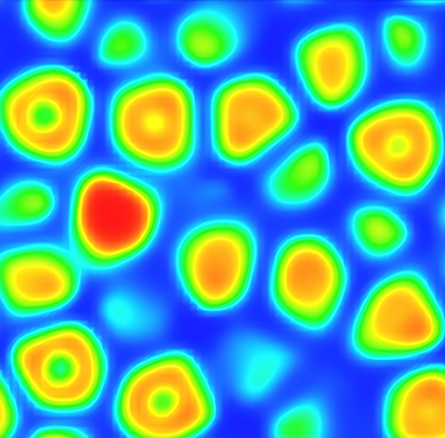}
\caption{After removing 2-coordinated O atom.}
\label{fig:315f}
\end{subfigure}
\caption{Electron localization function (ELF) slices for different models. (a), (c), (e) show ELF at the position of the oxygen atom before defect introduction. (b), (d), (f) show ELF near the defect site after vacancy formation.}
\label{fig:315}
\end{figure*}

The introduction of V$_O$s leads to increased electron tunneling probability in the oxide barrier layer and changes in chemical bonding. By analyzing the Electron Localization Function (ELF), we can identify regions of highly localized electrons and distinguish between metallic, covalent, and ionic bonds. We compared the ELF distributions around oxygen atoms in the defect-free model and those in the defect-introduced model. Figs.~\ref{fig:315a} and \ref{fig:315b} present the ELF maps before and after the introduction of a 4-coordinated oxygen vacancy, respectively. Prior to defect formation, pronounced regions of strong electron localization are observed between the oxygen atom and its neighboring Al atoms, distributed along the Al--O--Al bond directions. This indicates a high degree of electron localization in these regions, where localized electron pairs are stably formed. After the introduction of the oxygen vacancy, the ELF maps exhibit enhanced local asymmetry, accompanied by a pronounced reduction in the degree of electron localization. This change suggests that the removal of the oxygen atom disrupts the original bonding environment and weakens the localization of electrons in the vicinity of the defect. Figs.~\ref{fig:315c} and \ref{fig:315d} show the ELF around a 3-coordinated oxygen atom before and after vacancy formation, respectively. Upon vacancy formation, the electron localization at the defect site is significantly weakened, and the surrounding regions exhibit pronounced perturbations in the localization pattern. A relatively extended region with low ELF values emerges, indicating a reduction in local electron density and a decrease in electron localization. This behavior suggests that the original Al--O--Al bonding network is disrupted, leading to a redistribution of electronic states without the formation of new strong bonding configurations. Consequently, the electronic states become more delocalized, with a spatially diffuse charge density distribution.
Figs.~\ref{fig:315e} and \ref{fig:315f} display the ELF distribution before and after introducing a 2-coordinated V$_O$. Before defect formation, distinct high-ELF regions (highlighted in red) are observed between the oxygen atom and neighboring Al atoms, reflecting strong electron localization and pronounced Al--O bonding characteristics with a high degree of spatial symmetry. After introducing the V$_O$, the high-ELF region at the defect site does not exhibit a significant reduction, and a similar strong localized distribution is preserved. This indicates that the electron localization at the vacancy site is not directly destroyed by defect formation. However, substantial changes are observed in the ELF distribution in the surrounding regions. Some of the originally uniform high-ELF regions are weakened, while several intermediate-ELF regions become asymmetric, reflecting a vacancy-induced nonlocal reconstruction of the electronic structure. These results suggest that the oxygen vacancy strongly interacts with the surrounding bonding network, driving a rearrangement of the electronic states in its vicinity.

\begin{figure*}[htbp]
\centering
\begin{subfigure}{0.445\textwidth}
\centering
\includegraphics[width=\linewidth]{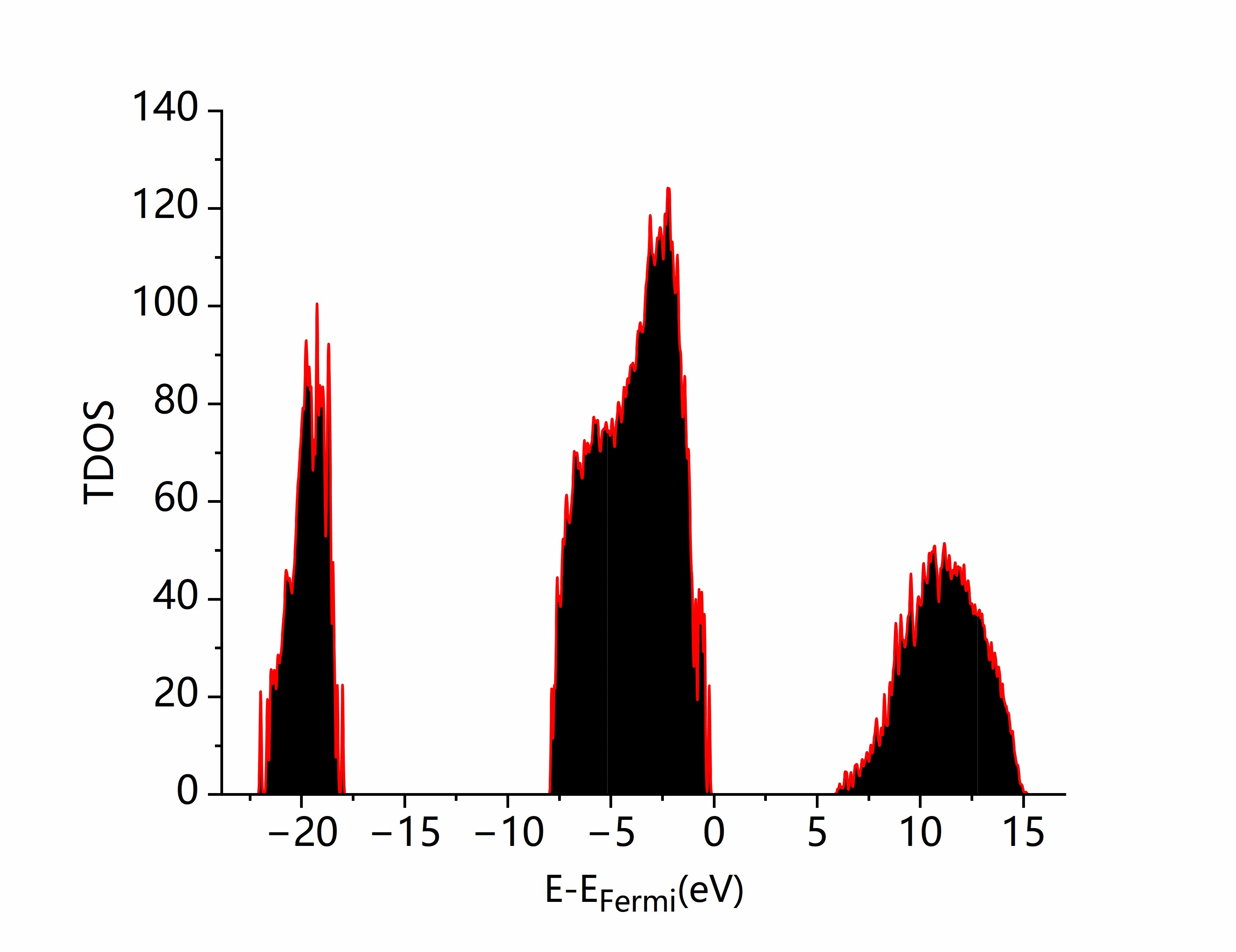}
\caption{Defect-free model.}
\label{fig:316a}
\end{subfigure}
\hfill
\begin{subfigure}{0.445\textwidth}
\centering
\includegraphics[width=\linewidth]{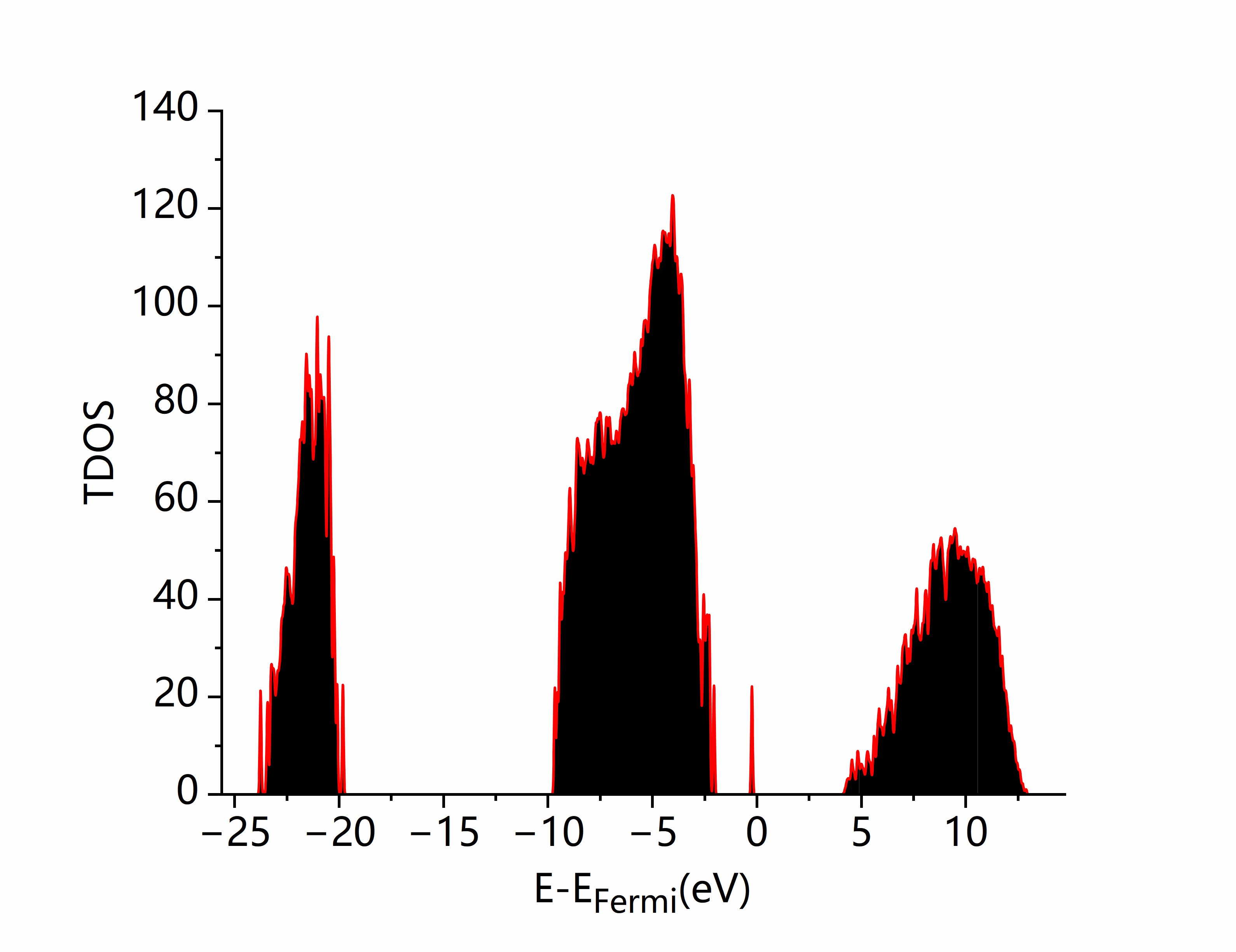}
\caption{4-coordinated V$_O$ model.}
\label{fig:316b}
\end{subfigure}
\vfill
\begin{subfigure}{0.445\textwidth}
\centering
\includegraphics[width=\linewidth]{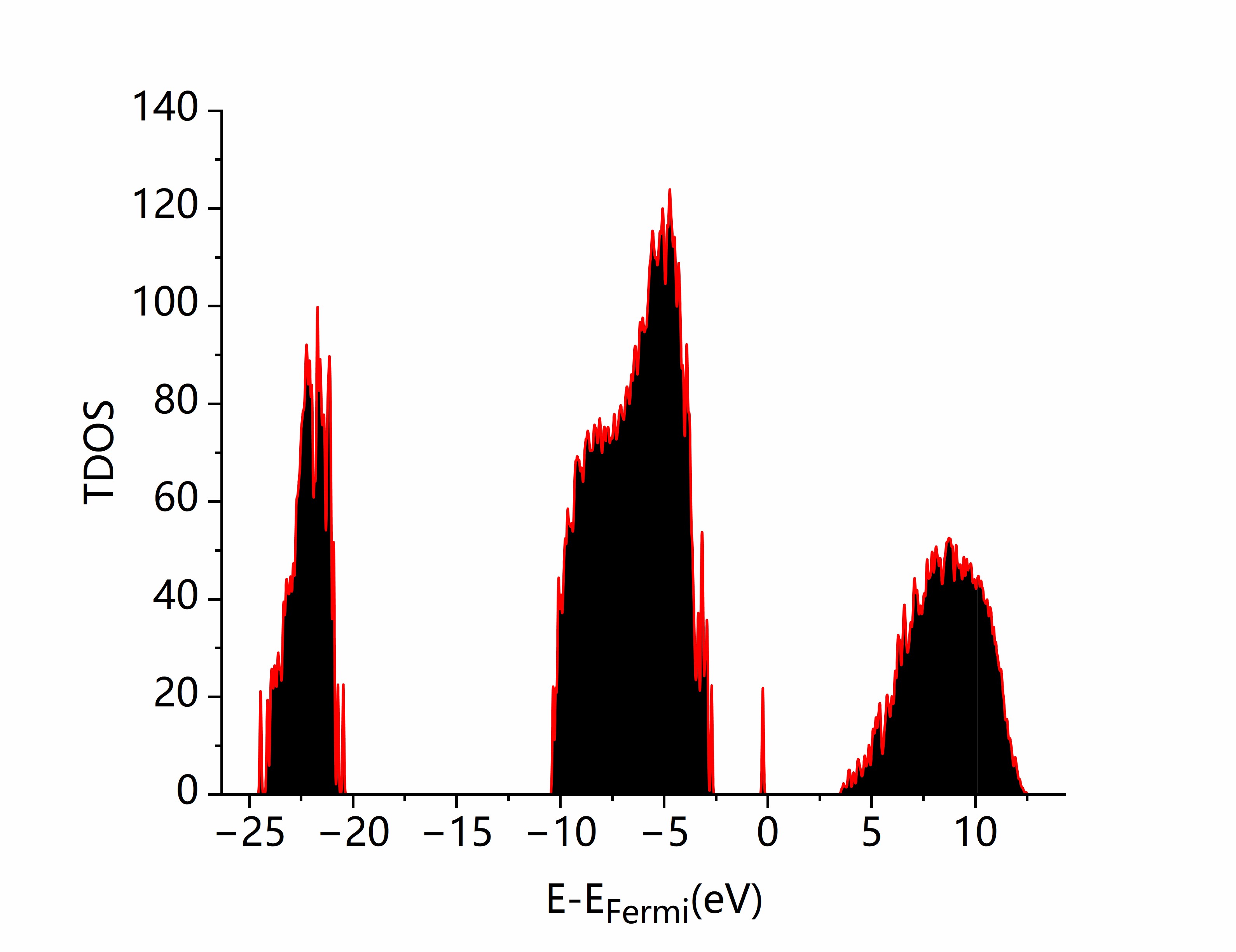}
\caption{3-coordinated V$_O$ model.}
\label{fig:316c}
\end{subfigure}
\hfill
\begin{subfigure}{0.445\textwidth}
\centering
\includegraphics[width=\linewidth]{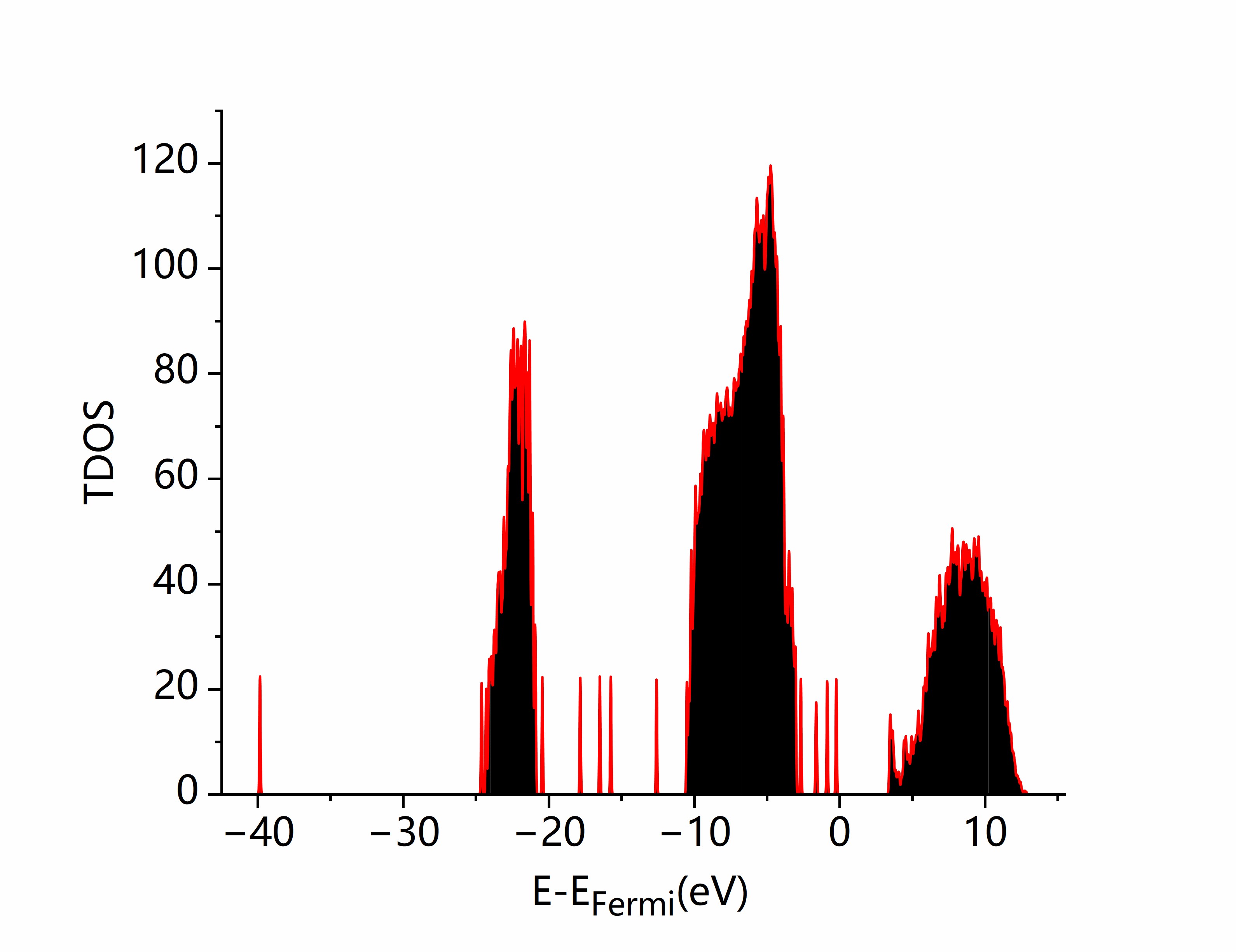}
\caption{2-coordinated V$_O$ model.}
\label{fig:316d}
\end{subfigure}
\caption{Total density of states (TDOS) for amorphous Al$_2$O$_3$ models with different V$_O$ coordination environments.}
\label{fig:316}
\end{figure*}

Comparing the total density of states (TDOS) between defect-free and defect-containing models (Fig.~\ref{fig:316}), we observe that the Fermi level is located near the middle of the band gap in the defect-free model. A finite density of states is present on the valence-band side, whereas the energy range from 0 to approximately +5 eV is nearly devoid of electronic states, with a pronounced increase in the density of states only appearing around +5 eV, close to the conduction band minimum. This behavior is consistent with the intrinsic characteristics of amorphous Al$_2$O$_3$ as a wide-bandgap, highly insulating dielectric material.
In contrast, upon the introduction of oxygen vacancies, emergent peaks in the electronic density of states are observed in the vicinity of the Fermi level (around 0 eV) for all defective models. This indicates that V$_O$s introduce new localized electronic states within the pristine band gap, thereby breaking the electronic-state symmetry of the defect-free system.
Although the positions and intensities of the defect-related peaks near 0 eV are comparable for the 4- and 3-coordinated V$_O$ models (Figs.~\ref{fig:316b} and \ref{fig:316c}), a more detailed defect-level analysis reveals notable differences. In the 4-coordinated model, the defect states are located closer to the valence band maximum, leading to a higher probability of electron trapping and a reduced ability to release free electrons for conduction. Consequently, the contribution of 4-coordinated V$_O$s to electrical conductivity enhancement is limited. In contrast, the defect states in the 3-coordinated model lie closer to the Fermi level, while the 2-coordinated model exhibits multiple defect states within the band gap (Fig.~\ref{fig:316d}). These additional states provide multiple channels for carrier generation and transport, facilitating carrier excitation and enhancing electrical conductivity.

Overall, V$_O$s introduce new localized states within the band gap of amorphous Al$_2$O$_3$, and the energy distribution and degree of spatial localization of these states are governed by the coordination environment of the vacancy. By modulating carrier generation and localization characteristics, these vacancy-induced states affect the electrical behavior of amorphous Al$_2$O$_3$.

\subsection{Influence of V$_O$ Concentration on Electrical Conductivity}

Under irradiation environments, the concentration of V$_O$ defects in amorphous Al$_2$O$_3$ increases with increasing irradiation dose, and the defect concentration directly influences the local electronic structure and macroscopic transport properties of the material. To systematically investigate how variations in V$_O$ concentration regulate the electrical conductivity of amorphous Al$_2$O$_3$, we randomly introduced 1, 2, 4, and 9 V$_O$s into the amorphous Al$_2$O$_3$ models to simulate irradiation-induced defects. The coordination environment of the vacancies was kept fixed to eliminate the influence of coordination-related variations on the transport properties, thereby enabling an isolated assessment of the effect of vacancy concentration.

\begin{figure}[htbp]
\centering
\includegraphics[scale=0.65]{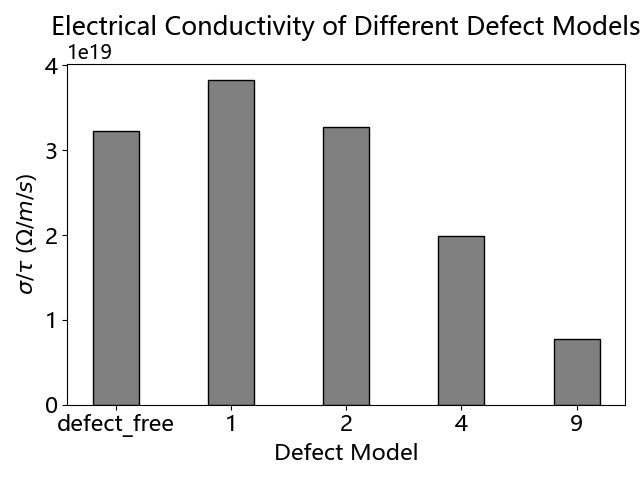}
\caption{Electrical conductivity ( $\sigma/\tau$) comparison of amorphous Al$_2$O$_3$ models with different numbers of V$_O$s (N = 0, 1, 2, 4, 9).}
\label{fig:321}
\end{figure}

\begin{figure*}[htbp]
\centering
\begin{subfigure}{0.48\textwidth}
\centering
\includegraphics[width=\linewidth]{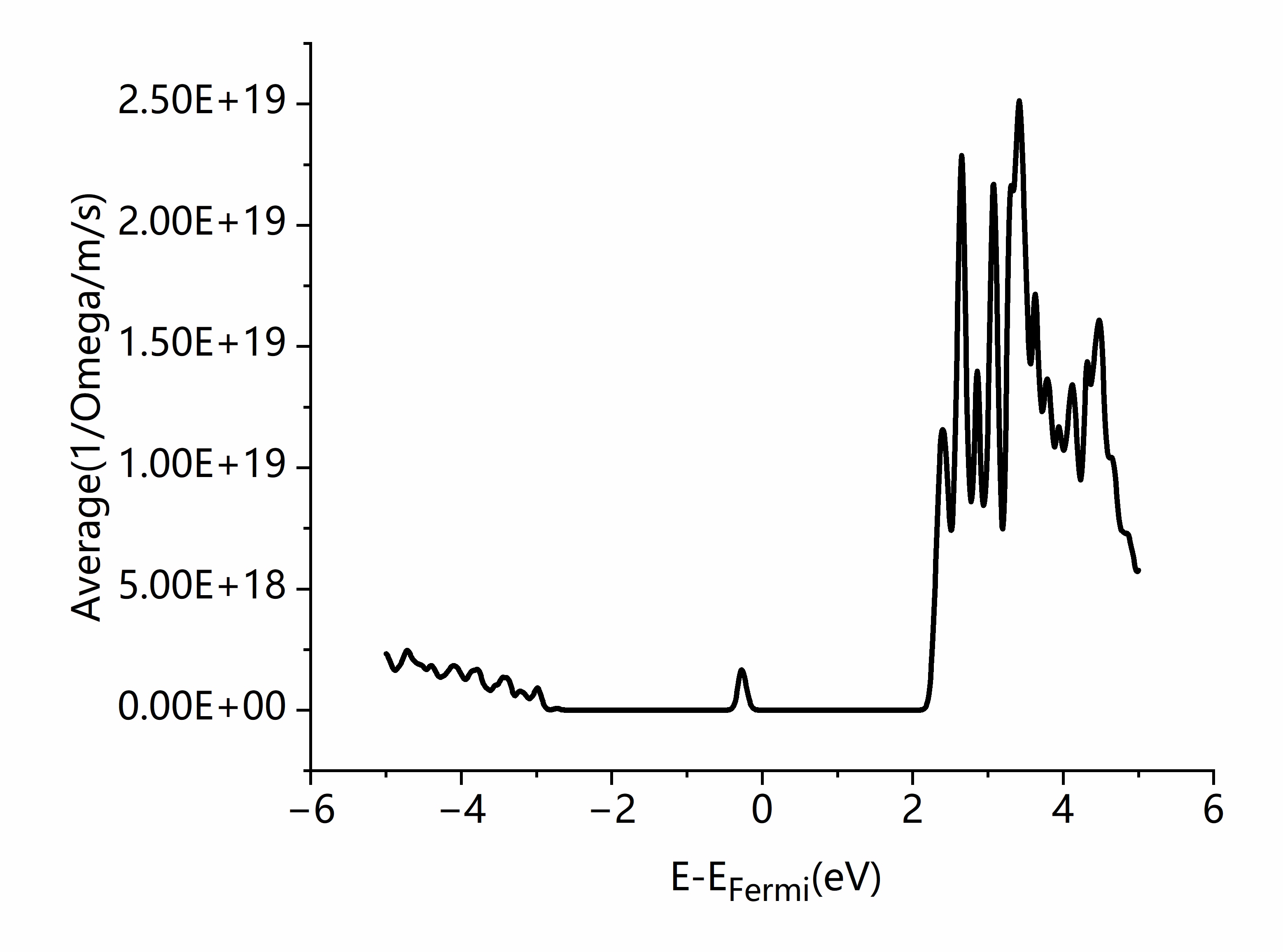}
\caption{Model with one V$_O$ (N = 1).}
\label{fig:322a}
\end{subfigure}
\hfill
\begin{subfigure}{0.48\textwidth}
\centering
\includegraphics[width=\linewidth]{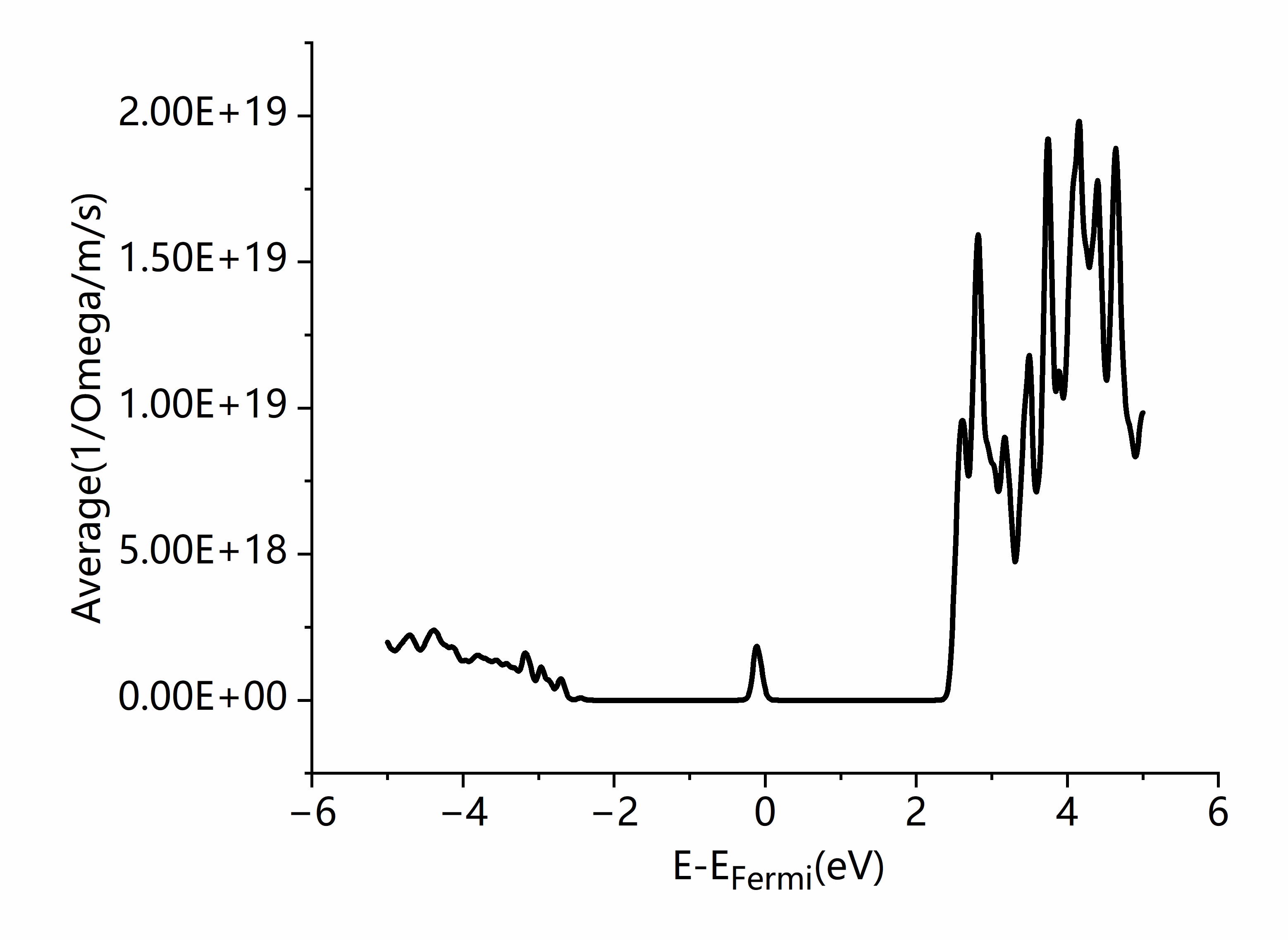}
\caption{Model with two V$_O$s (N = 2).}
\label{fig:322b}
\end{subfigure}
\vfill
\begin{subfigure}{0.48\textwidth}
\centering
\includegraphics[width=\linewidth]{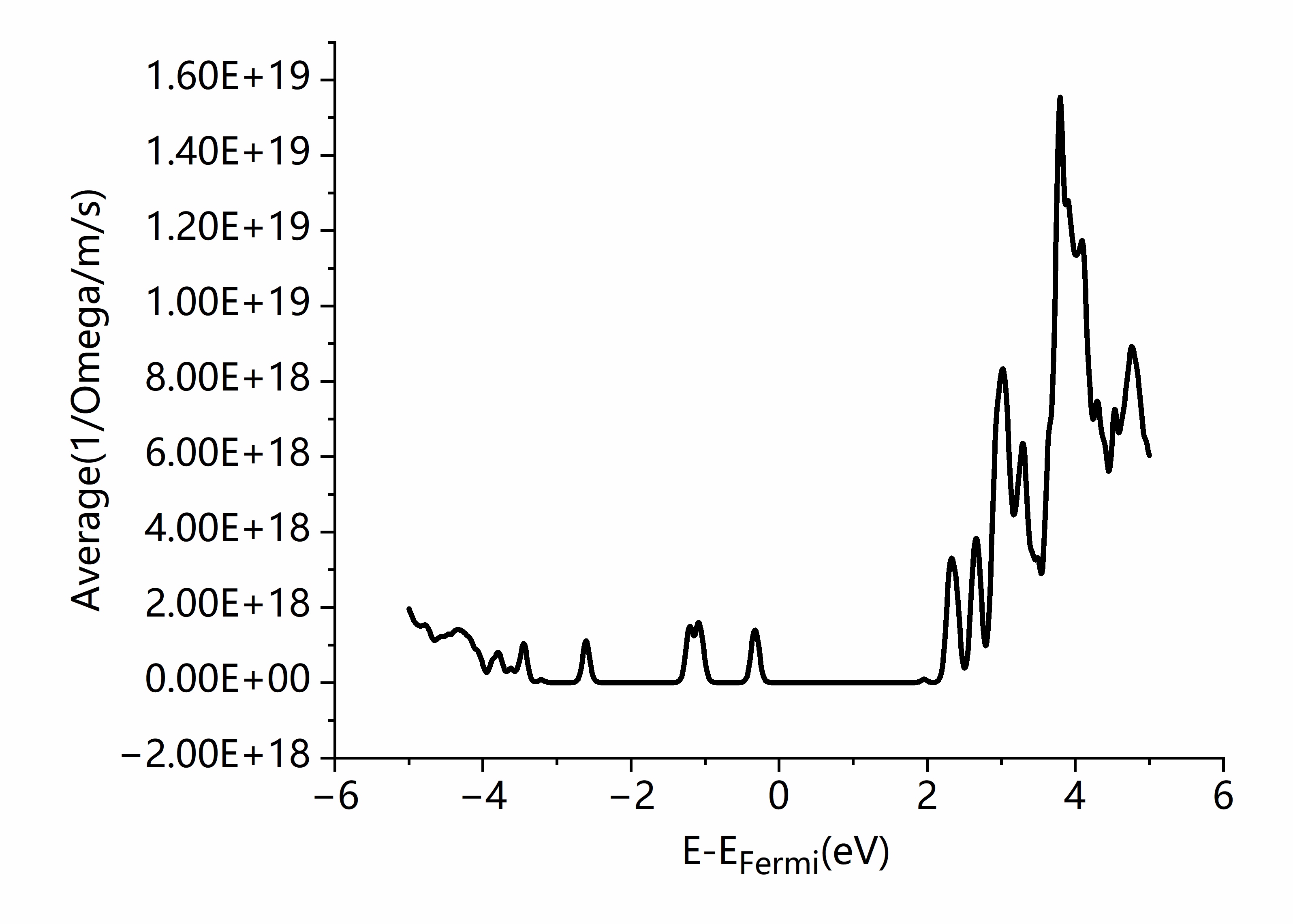}
\caption{Model with four V$_O$s (N = 4).}
\label{fig:322c}
\end{subfigure}
\hfill
\begin{subfigure}{0.48\textwidth}
\centering
\includegraphics[width=\linewidth]{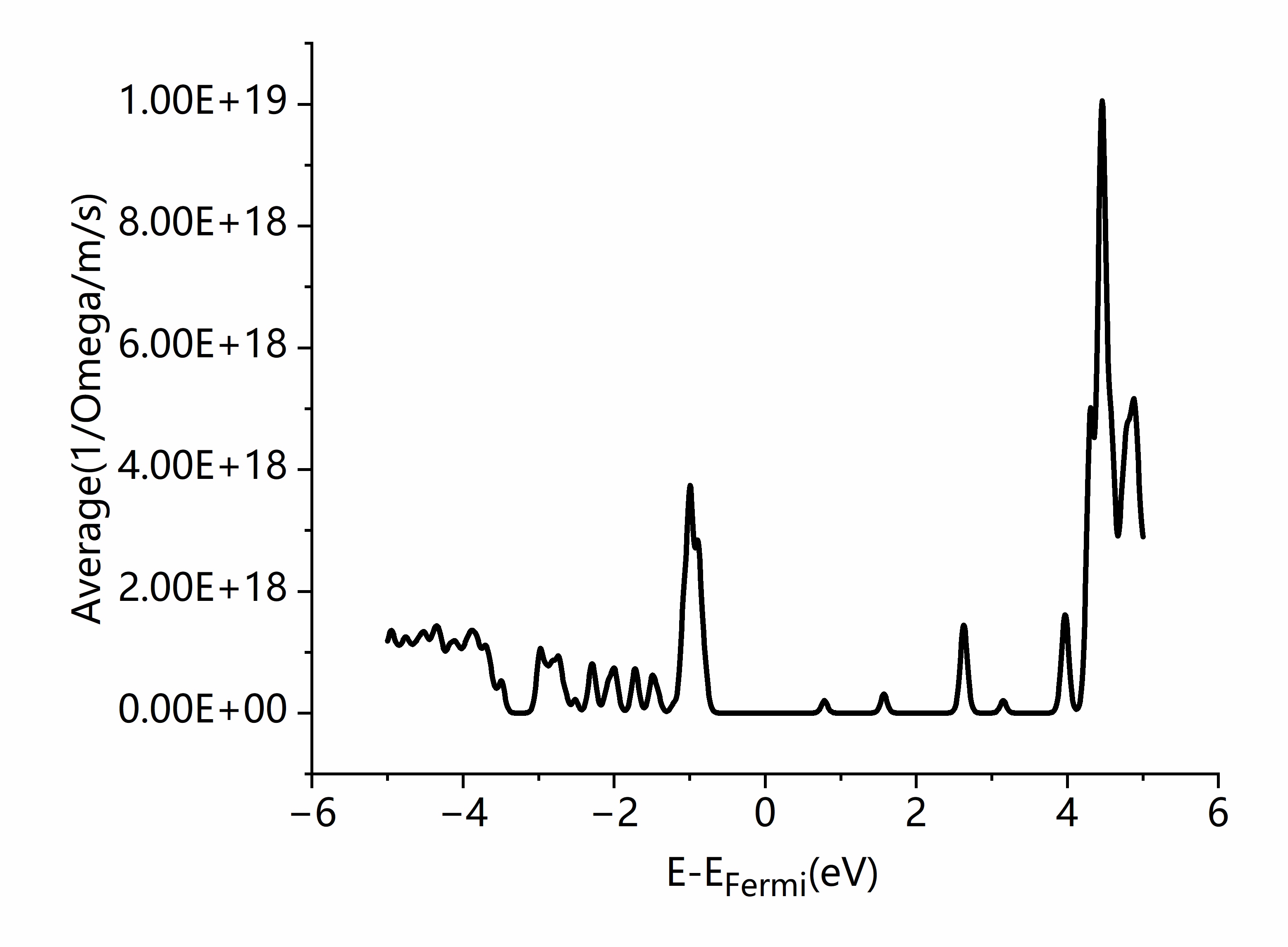}
\caption{Model with nine V$_O$s (N = 9).}
\label{fig:322d}
\end{subfigure}
\caption{Electrical conductivity spectra ( $\sigma/\tau$) for amorphous Al$_2$O$_3$ models with different numbers of V$_O$s.}
\label{fig:322}
\end{figure*}

\begin{figure}[htbp]
\centering
\includegraphics[scale=0.65]{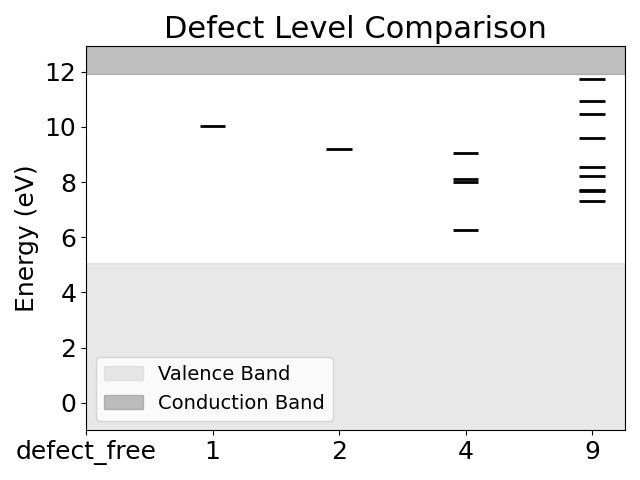}
\caption{Defect level distributions for amorphous Al$_2$O$_3$ models with different numbers of V$_O$s.}
\label{fig:323}
\end{figure}

\begin{figure*}[htbp]
\centering
\begin{subfigure}{0.48\textwidth}
\centering
\includegraphics[width=\linewidth]{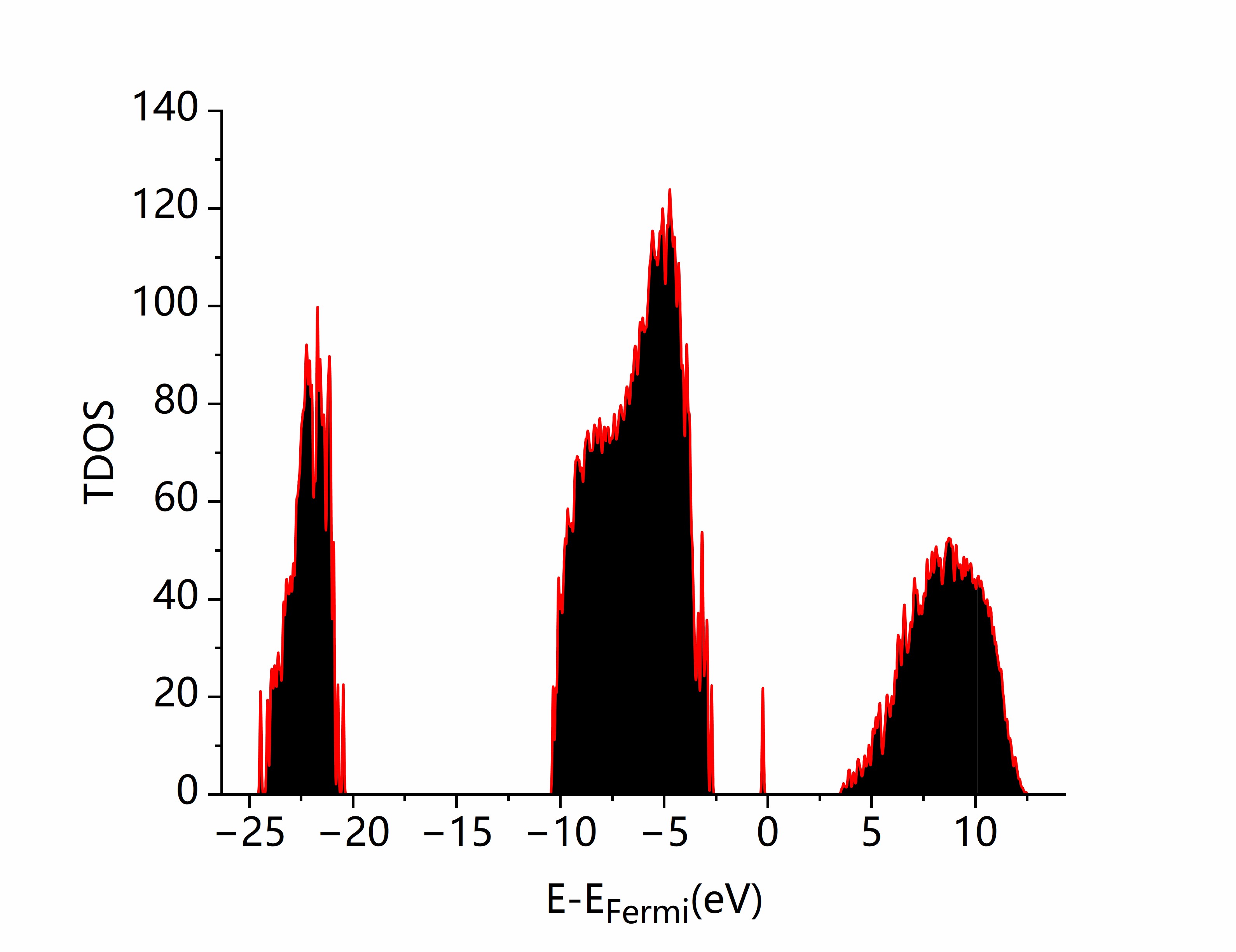}
\caption{N = 1}
\label{fig:324a}
\end{subfigure}
\hfill
\begin{subfigure}{0.48\textwidth}
\centering
\includegraphics[width=\linewidth]{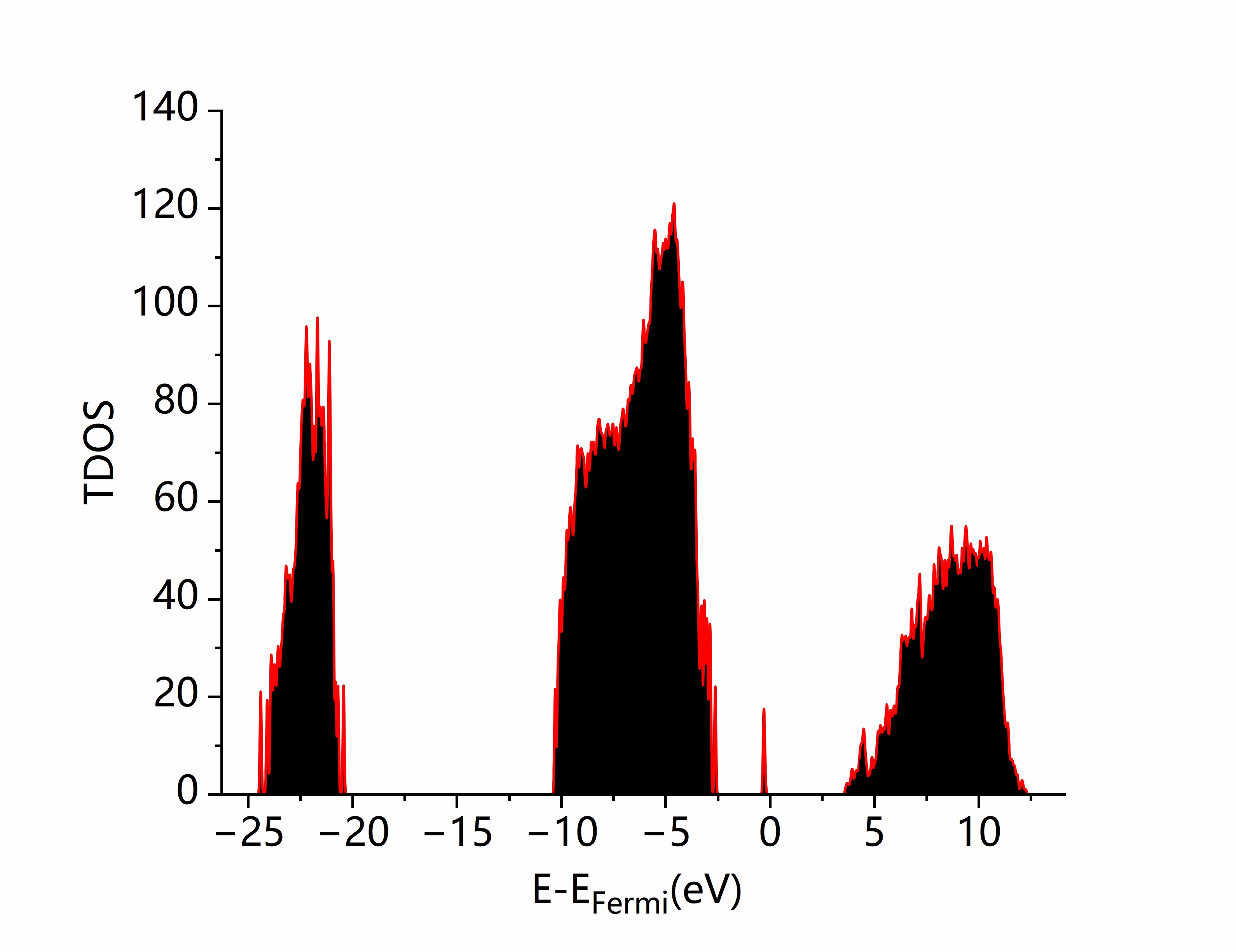}
\caption{N = 2}
\label{fig:324b}
\end{subfigure}
\vfill
\begin{subfigure}{0.48\textwidth}
\centering
\includegraphics[width=\linewidth]{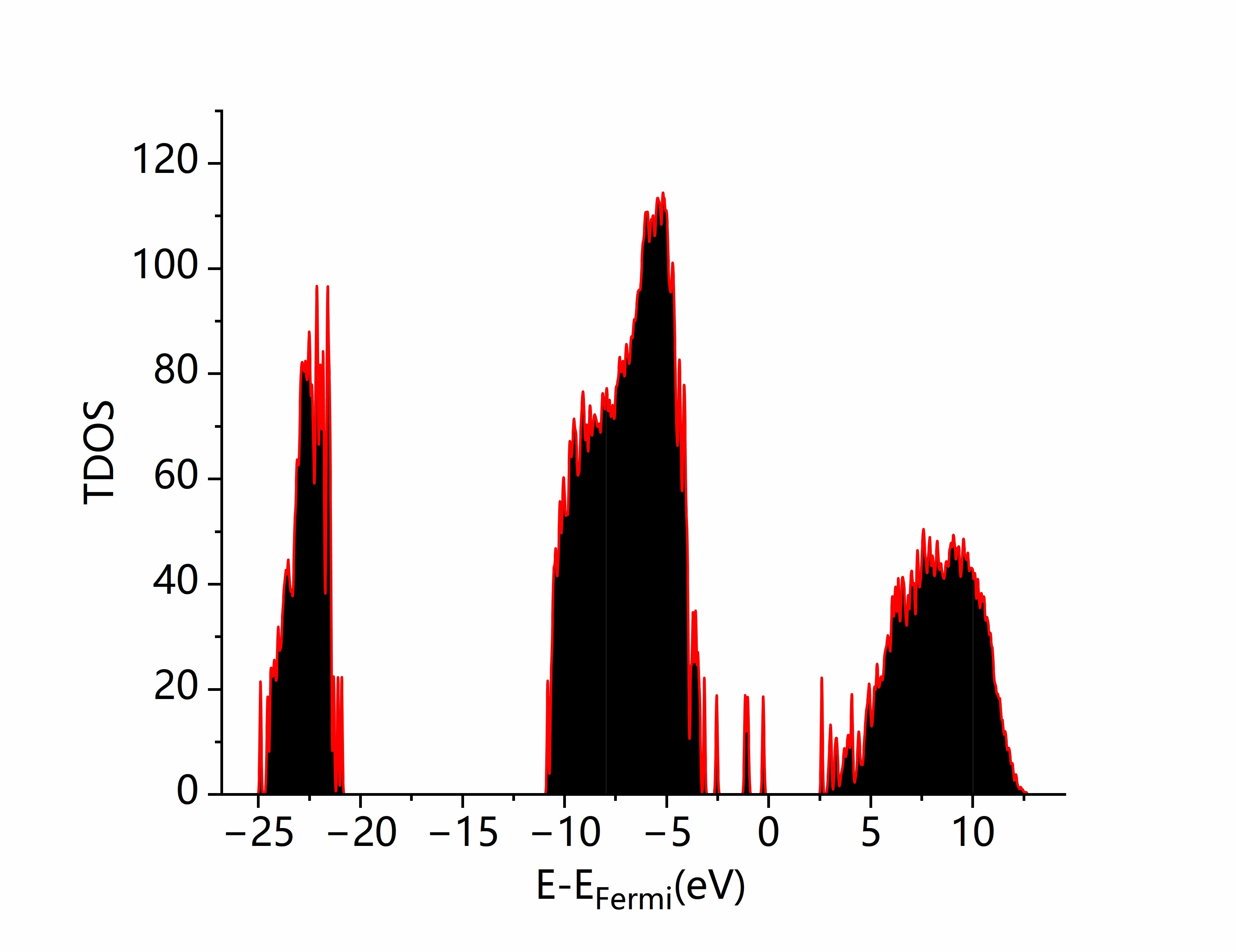}
\caption{N = 4}
\label{fig:324c}
\end{subfigure}
\hfill
\begin{subfigure}{0.48\textwidth}
\centering
\includegraphics[width=\linewidth]{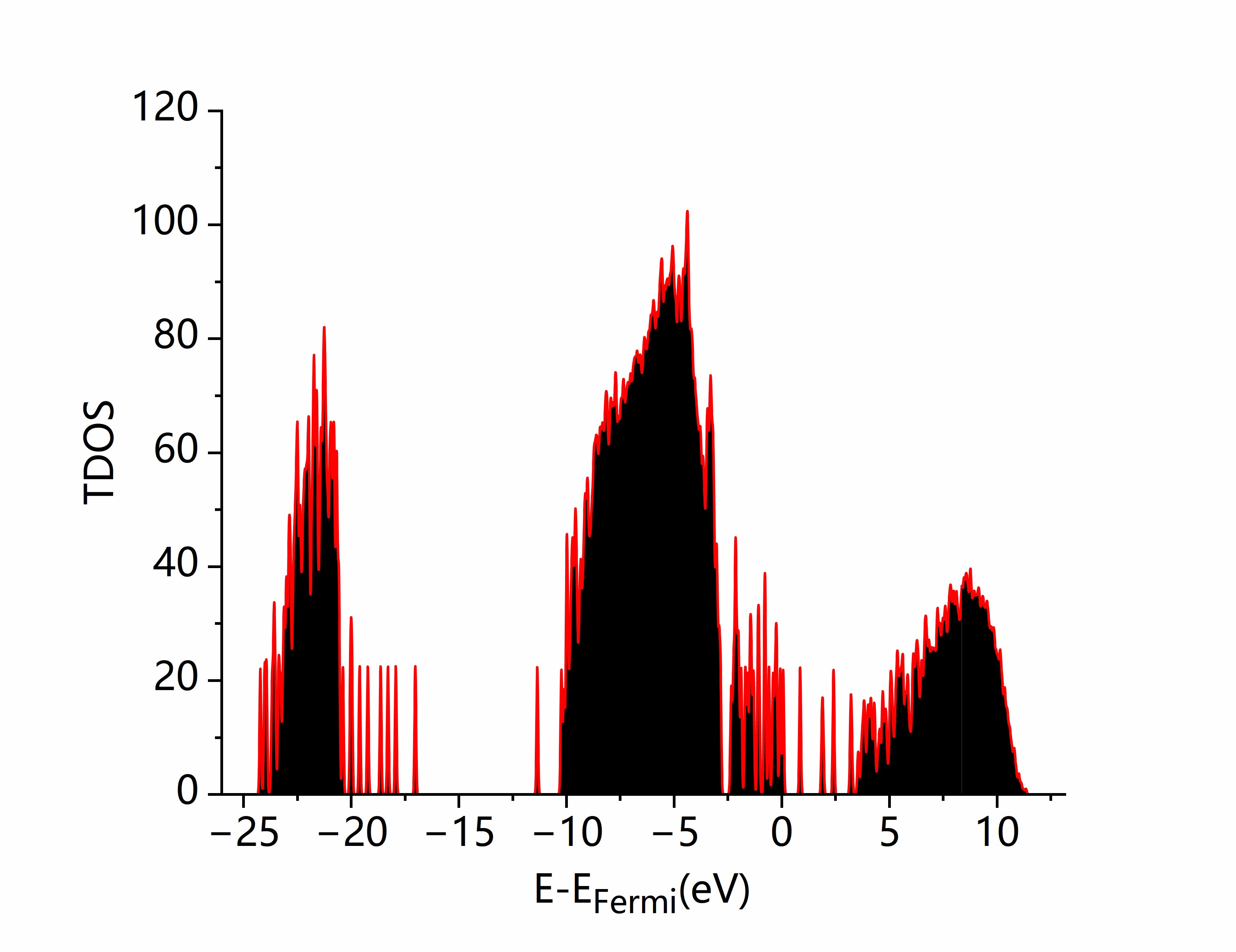}
\caption{N = 9}
\label{fig:324d}
\end{subfigure}
\caption{Total density of states for amorphous Al$_2$O$_3$ models with different numbers of V$_O$s.}
\label{fig:324}
\end{figure*}

As shown in Fig.~\ref{fig:321}, the introduction of a single V$_O$ leads to an enhancement of the electrical conductivity compared with the defect-free model, indicating that an isolated V$_O$ can introduce shallow donor states within the band gap and increase the concentration of free charge carriers. By comparing the conductivity spectra of different models (Fig.~\ref{fig:322}), we find that the introduction of a single V$_O$ gives rise to a weak peak near the Fermi level (0--1 eV). Consistent with the density-of-states results (Fig.~\ref{fig:324}), a newly emerged electronic state appears near the Fermi level, in agreement with the defect-level analysis (Fig.~\ref{fig:323}). This defect state lies close to the conduction band minimum and can be identified as a typical shallow donor level. Such shallow donor states are capable of releasing electrons into the conduction band, thereby increasing the free carrier concentration and enhancing the probability of electronic transitions. These results indicate that, at low defect concentrations, V$_O$s primarily act as electron donors, providing additional conductive pathways for the material.

When two V$_O$s are introduced, the overall conductivity exhibits little change and remains comparable to that of the defect-free model, indicating that the transport properties are only weakly affected. Although a peak near the Fermi level can still be observed in the conductivity spectrum, the presence of two V$_O$s induces only limited local structural perturbations and does not yet lead to the formation of deep defect levels or vacancy clustering.
With further increases in vacancy concentration, the electrical conductivity exhibits a monotonic evolution that is consistent with trends observed in experimental simulations. Upon introducing four V$_O$s, the conductivity decreases markedly, showing a behavior that is fundamentally different from that observed at low defect concentrations. This suggests that higher concentrations of V$_O$s tend to introduce deeper donor states, which are less effective at supplying free carriers and may even act as carrier traps, thereby hindering electronic transport. The conductivity spectrum shows a clear overall change, with multiple peaks appearing near the Fermi level and a pronounced reduction in conductivity in the low-energy region. The associated defect levels are predominantly located closer to the valence band (Fig.~\ref{fig:323}). In addition, interactions between multiple vacancies may lead to defect-defect coupling or the formation of localized vacancy clusters, resulting in repeated carrier scattering within confined regions. Such effects suppress effective conductive channels and impede carrier migration, ultimately leading to degraded transport performance. Therefore, although increasing V$_O$ concentration can introduce more electronic states within the band gap, these states do not necessarily contribute to conduction.

The density-of-states (DOS) results for all models are shown in Fig.~\ref{fig:324}. With increasing V$_O$ concentration, the number of electronic states within the band gap increases and the effective band gap narrows. In principle, such changes could facilitate electronic excitation and enhance electrical conductivity. However, the calculated conductivity instead decreases, indicating that irradiation-induced high concentrations of V$_O$s in amorphous Al$_2$O$_3$ possess a strong electron-trapping capability. This behavior highlights the distinct influence of high-density V$_O$s on the band structure and transport properties of amorphous materials.

In amorphous systems, the electrical properties exhibit a nonlinear dependence on V$_O$ concentration. At low defect concentrations, shallow donor states introduced by V$_O$s enhance electrical conductivity. In contrast, high concentrations of V$_O$s do not further improve conductivity but instead promote carrier trapping and suppress electron tunneling. As a result, the effective tunneling-blocked junction area increases, leading to a degradation of the overall transport performance.

\section{Impact of Oxygen Vacancies on Josephson Junction Decoherence Mechanisms}\label{sec:ov}
\subsection{Decoherence Mechanisms}

Since charge transport is intrinsically linked to the critical current of Josephson junctions, conductivity variations induced by V$_O$s can introduce additional low-frequency noise sources. In particular, such defect-induced fluctuations can give rise to critical current noise with a pronounced $1/f$ characteristic near the operating frequencies of superconducting qubits, directly affecting the energy level splitting and phase stability. Therefore, we discuss the mechanisms by which V$_O$ defects influence qubit decoherence from both theoretical and numerical perspectives.

According to Van Harlingen et al.~\cite{vandecoherence24}, the low-frequency $1/f$ noise spectral density of the critical current in Josephson junctions is correlated with the decoherence time of superconducting qubits, as shown in Eq.~\ref{Eq:critical}:
\begin{equation}
\label{Eq:critical}
T_\phi \propto \frac{I_0}{\Omega \Lambda \sqrt{S_{I_0}(1\,\mathrm{Hz})}},
\end{equation}
where $\Omega$ is the tunneling frequency, and $\Omega/2\pi$ is the energy level splitting frequency. For a superconducting qubit, the energy difference between the two quantum states is $\hbar \Omega$. 

The parameter $\Lambda$ represents the relative variation in energy level splitting caused by the relative change of the critical current,
\begin{equation}
\label{Eq:2}
\Lambda = \left|\frac{I_0}{\Omega} \frac{d \Omega}{d I_0}\right|,
\end{equation}
and indicates the sensitivity of the qubit to fluctuations in critical current. The larger $\Lambda$ is, the more sensitive the qubit is to critical current fluctuations and the more easily it is affected by decoherence.

The spectral density of critical current noise at 1~Hz can be approximated as
\begin{equation}
\label{Eq:3}
S_{I_{0}}^{1/2}(1\,\mathrm{Hz}) \approx 144 \left(\frac{I_{0}}{\mu\mathrm{A}}\right) \left(\frac{\mu\mathrm{m}^{2}}{A}\right)^{1/2} \mathrm{pA}/\sqrt{\mathrm{Hz}},
\end{equation}
where $A$ is the junction area. The decoherence time can be estimated through this spectral density.

The critical current and the junction resistance are linked by the relation
\begin{equation}
\label{Eq:4}
I_{0} = \frac{\pi \Delta}{2 e R_{N}},
\end{equation}
where $\Delta$ is the superconducting energy gap, and $R_{N}$ is the junction resistance.

According to Eq.~\ref{Eq:4}, conductivity fluctuations lead to variations in the junction resistance, thereby influencing the critical current. Such fluctuations induced by V$_O$s can be regarded as the superposition of multiple random telegraph signals~\cite{vandecoherence24}, forming $1/f$ low-frequency noise that modulates the qubit energy level splitting frequency and consequently reduces the coherence time.
Single charge traps generate telegraph noise characterized by two-state lifetimes (occupied/unoccupied), contributing Lorentzian noise peaks with an effective correlation time
\begin{equation}
\label{Eq:5}
\tau_{\mathrm{eff}} = \left(\frac{1}{\tau_t} + \frac{1}{\tau_u}\right)^{-1}.
\end{equation}

For a junction of area $A$, the critical current change $\Delta I_0$ is proportional to the effective blocked tunneling area $DA$ by trapped charges,
\begin{equation}
\label{Eq:6}
\Delta I_{0} = \left(\frac{DA}{A}\right) I_{0}.
\end{equation}
If a charge is trapped, the effective junction area blocked for tunneling increases.

The spectral density of the critical current noise induced by a single trap is proportional to $(\Delta I_0)^2$. $S_{I_0}$ is proportional to $I_0^2/A$. 
As the irradiation dose increases, multiple V$_O$ traps emerge within the oxide barrier, and the superposition of their Lorentzian spectra forms a $1/f$-like noise spectrum. The spectral density of $N$ identical traps is proportional to $N(\Delta I_0)^2$:
\begin{equation}
\label{Eq:7}
N(\Delta I_0)^2 = nA \left(\frac{DA}{A}\right)^2 I_0^2,
\end{equation}
where $n$ denotes the trap density per unit area. Consequently, with increasing irradiation dose, the concentration of V$_O$ defects increases, leading to an enhancement of critical current noise. The enhanced noise spectral density is associated with a shortened decoherence time. The temperature dependence of $1/f$ critical current noise remains inconclusive. Since charge traps predominantly operate via tunneling at low temperatures, the temperature dependence is expected to be weak, and thus the described noise spectral density relation is applicable in our study.

Using the time-domain scan-averaging method and experimental parameters ($I_0 =1$~mA, $A=0.01$~mm$^2$, $\Lambda =100$, $\Omega/2\pi=1$~GHz, $T =100$~mK), decoherence times are estimated between 0.8 ms and 12 ms (Eq.~\ref{Eq:8}), depending on junction area:
\begin{equation}
\label{Eq:8}
t_f^B(\mathrm{ms}) \approx 15 \times A^{1/2}(\mu\mathrm{m}) \times \frac{1}{\Lambda} \times \frac{1}{\Omega(\mathrm{GHz})} \times \frac{T}{4.2\,\mathrm{K}}.
\end{equation}

\subsection{Estimation of Decoherence Times for Different Models}

To quantitatively evaluate the impact of different types of V$_O$ defects on the decoherence performance of superconducting qubits, we estimated the decoherence times for each model based on the critical current noise mechanism proposed by Van Harlingen et al.~\cite{vandecoherence24}, in conjunction with the conductivity variations obtained in Sec.~\ref{sec:results}.

Since the exact relaxation time in amorphous Al$_2$O$_3$ is difficult to determine, this work focuses on the relative variation of $\sigma/\tau$ among different vacancy configurations, which reflects the transport changes induced by V$_O$s. As the relaxation time $\tau$ is expected to vary weakly among these configurations, it cancels out when considering the relative fluctuations of the critical current; therefore, the decoherence analysis can be performed directly using the calculated $\sigma/\tau$ values.

According to Eq.~\ref{Eq:4}, assuming a fixed superconducting energy gap, the critical current $I_0$ is proportional to the electrical conductivity:
\begin{equation}
\label{Eq:9}
R_{N} \propto \frac{1}{\sigma}, \qquad I_0 \propto \sigma \cdot \Delta.
\end{equation}
When the conductivity changes, it induces a slight perturbation in the resistance:
\begin{equation}
\label{Eq:10}
\frac{\Delta I_{0}}{I_{0}} = -\frac{\Delta R_{N}}{R_{N}} = \frac{\Delta \sigma}{\sigma},
\end{equation}
so that
\begin{equation}
\label{Eq:11}
\Delta I_{0} \propto \frac{\Delta \sigma}{\sigma}.
\end{equation}
Therefore, variations in conductivity give rise to corresponding changes in the critical current. As a result, the noise spectral density can be written as
\begin{equation}
\label{Eq:12}
S_{I_0}(f) \propto N(\Delta I_0)^2 \propto N\left(\frac{\Delta\sigma}{\sigma}\right)^2,
\end{equation}
and the decoherence time becomes
\begin{equation}
\label{Eq:13}
T_{\phi} \propto \frac{1}{S_{I_0}(f)} \propto \frac{1}{N\left(\Delta\sigma / \sigma\right)^{2}}.
\end{equation}
To avoid divergence and to compare relative decoherence times, we employ a phenomenological normalization:
\begin{equation}
\label{Eq:14}
T_{\phi} = \frac{T_{\phi,0}}{1 + N \left(\Delta\sigma / \sigma_{0}\right)^{2}},
\end{equation}
where $\sigma_{0}$ is the conductivity of the defect-free model and $T_{\phi,0}$ is its decoherence time. This expression captures the qualitative trend that larger conductivity fluctuations and higher defect densities reduce coherence.

Based on the derived theoretical framework, the decoherence times of superconducting qubits in the presence of oxygen-vacancy defects can be estimated. Using experimentally relevant parameters and assuming a decoherence time of 1 ms for the defect-free model, the decoherence times for each model are obtained and summarized in Tables~\ref{Tab:2} and \ref{Tab:3}. The results are consistent with theoretical expectations and exhibit trends in agreement with experimental observations.

\renewcommand{\arraystretch}{1.5} 
\begin{table}[htbp]
\centering
\caption{Conductivity $\sigma/\tau$, relative fluctuation $\Delta\sigma/\sigma_0$, and estimated decoherence time $T_\phi$ for V$_O$ defect models with different coordination numbers. $\sigma_0 = 3.23 \times 10^{19}~\Omega^{-1}\mathrm{m}^{-1}\mathrm{s}^{-1}$ is the conductivity of the defect-free model.}
\begin{tabular}{lccc}
\hline
\textbf{Model} & $\sigma/\tau$ ($\Omega^{-1}\mathrm{m}^{-1}\mathrm{s}^{-1}$) & $\Delta\sigma / \sigma_0$ & $T_\phi$ (ms) \\
\hline
Defect-free   & $3.23 \times 10^{19}$ & 0                & 1.000          \\
4-coordinated & $2.48 \times 10^{19}$ & $-$0.232         & 0.949          \\
3-coordinated & $3.83 \times 10^{19}$ & $+$0.186         & 0.967          \\
2-coordinated & $3.51 \times 10^{19}$ & $+$0.086         & 0.993          \\
\hline
\end{tabular}
\label{Tab:2}
\end{table}

\renewcommand{\arraystretch}{1.5} 
\begin{table}[htbp]
\centering
\caption{Conductivity $\sigma/\tau$, relative fluctuation $\Delta\sigma/\sigma_0$, and estimated decoherence time $T_\phi$ for V$_O$ defect models with different numbers of vacancies.}
\begin{tabular}{lccc}
\hline
\textbf{Number of V$_O$s} & $\sigma/\tau$ ($\Omega^{-1}\mathrm{m}^{-1}\mathrm{s}^{-1}$) & $\Delta\sigma / \sigma_0$ & $T_\phi$ (ms) \\
\hline
0 & $3.23 \times 10^{19}$ & 0       & 1.000          \\
1 & $3.83 \times 10^{19}$ & $+$0.186 & 0.967          \\
2 & $3.27 \times 10^{19}$ & $+$0.012 & 0.999          \\
4 & $1.99 \times 10^{19}$ & $-$0.380 & 0.634          \\
9 & $7.78 \times 10^{18}$ & $-$1.400 & 0.053          \\
\hline
\end{tabular}
\label{Tab:3}
\end{table}

As can be seen from the data summarized in Tables~\ref{Tab:2} and \ref{Tab:3}, both the coordination environment and the concentration of defects have a significant impact on the decoherence time of superconducting qubits. In amorphous Al$_2$O$_3$, the introduction of conventional 4-coordinated V$_O$s maintains the overall insulating character of the material; however, it leads to the largest conductivity fluctuations, resulting in the shortest decoherence time. In contrast, 3- and 2-coordinated V$_O$s increase the conductivity while inducing relatively smaller conductivity fluctuations, and consequently cause a less pronounced reduction in the decoherence time compared with the 4-coordinated case. As the number of V$_O$s increases, except for anomalous behavior observed at low defect concentrations, the conductivity exhibits a decreasing trend, while the amplitude of conductivity fluctuations increases with defect concentration. Correspondingly, the decoherence time decreases rapidly as the vacancy concentration increases.
These results indicate that, under irradiation conditions, a higher density of oxygen-vacancy defects in superconducting qubits leads to shorter decoherence times and more severe degradation of device performance. This trend is consistent with experimental observations.

\subsection{Impact on Rabi Oscillations}\label{sec:rabi}

To illustrate how the estimated decoherence times affect the coherent dynamics of a superconducting qubit, we simulate Rabi oscillations---the coherent oscillations of the qubit state under resonant microwave driving. The excited-state probability of a qubit undergoing Rabi oscillations with decoherence time $T_2$ is given by
\begin{equation}
P_{\mathrm{ex}}(t) = \frac{1}{2}\Bigl[1 - e^{-t/T_2} \cos(\Omega_R t)\Bigr],
\end{equation}
where $\Omega_R$ is the Rabi frequency. For a typical Rabi frequency $\Omega_R/2\pi = 2\;\mathrm{MHz}$, we plot the envelope of $P_{\mathrm{ex}}(t)$ (i.e., the curves $ \frac{1}{2}[1 \pm e^{-t/T_2}] $) for the decoherence times obtained from our models (Tables~\ref{Tab:2} and \ref{Tab:3}). The time window extends to $200\;\mu\mathrm{s}$, corresponding to roughly $4 T_2$ for the shortest coherence time ($T_2 = 0.053\;\mathrm{ms}$) and $0.2 T_2$ for the longest ($T_2 = 1.000\;\mathrm{ms}$).

\begin{figure*}[htbp]
\centering
\includegraphics[width=0.9\linewidth]{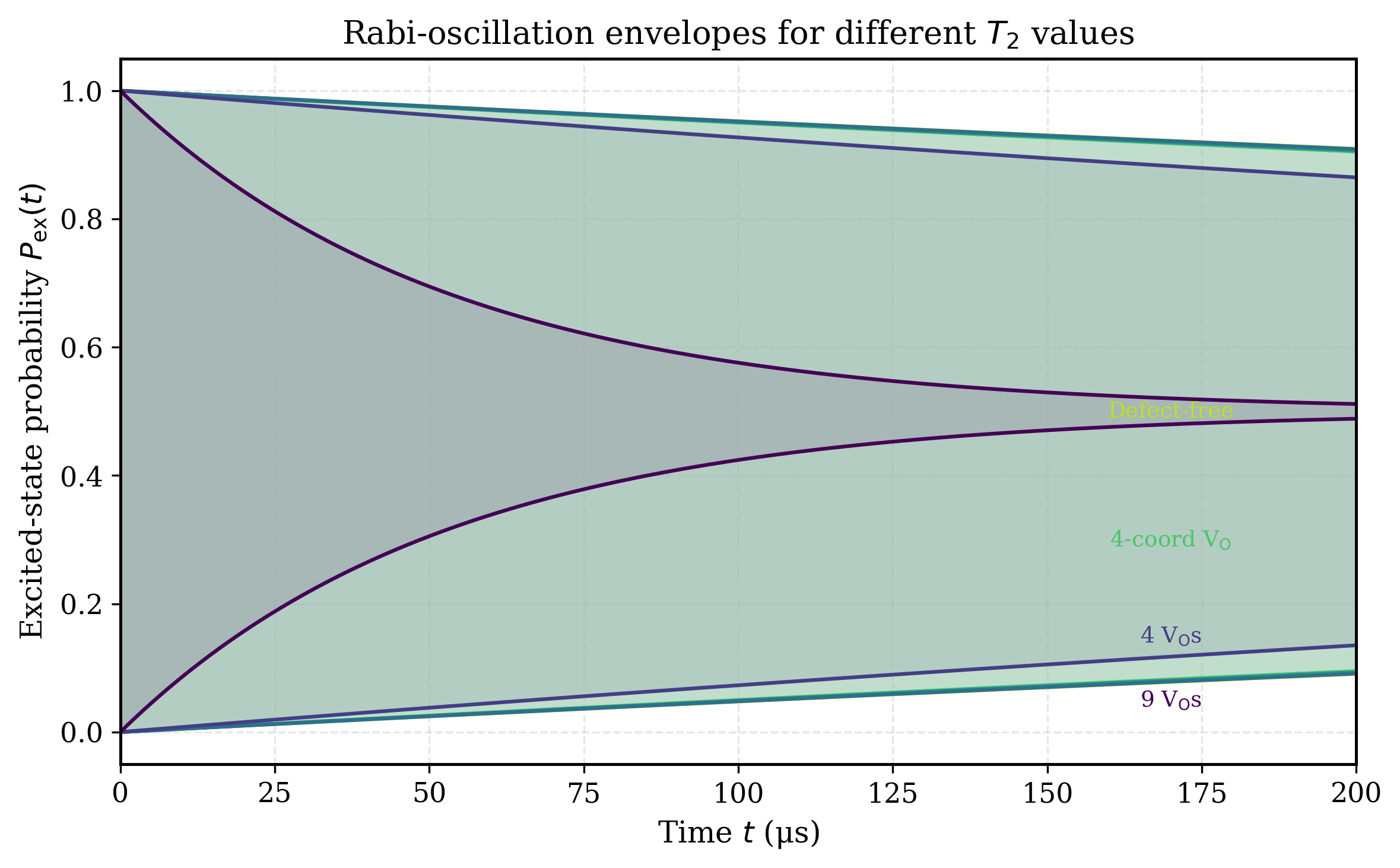}
\caption{Envelope of Rabi oscillations for different decoherence times $T_2$ estimated from the V$_O$ defect models. The shaded region between the upper and lower envelopes ($\frac{1}{2}[1 \pm e^{-t/T_2}]$) indicates the range of the oscillating probability $P_{\mathrm{ex}}(t)$. The defect‑free case ($T_2 = 1.000\;\mathrm{ms}$) shows almost no damping over the simulated $200\;\mu\mathrm{s}$ window, while the model with nine V$_O$s ($T_2 = 0.053\;\mathrm{ms}$) exhibits nearly complete damping within $\sim 50\;\mu\mathrm{s}$. Intermediate vacancy configurations (4‑coordinated V$_O$, $T_2 = 0.949\;\mathrm{ms}$; 4~V$_O$s, $T_2 = 0.634\;\mathrm{ms}$) display progressively stronger damping. The plot directly visualizes how oxygen‑vacancy‑induced conductivity fluctuations degrade qubit coherence and limit the number of coherent gate operations that can be performed.}
\label{fig:rabi}
\end{figure*}

Figure~\ref{fig:rabi} clearly demonstrates that even modest reductions in $T_2$ (e.g., $0.949\;\mathrm{ms}$ for a 4‑coordinated vacancy) cause noticeable narrowing of the envelope, corresponding to faster decay of the oscillation amplitude. When the vacancy concentration increases (e.g., nine V$_O$s, $T_2 = 0.053\;\mathrm{ms}$), the envelope collapses within $\sim 50\;\mu\mathrm{s}$, meaning the Rabi oscillations are almost completely suppressed after a few tens of microseconds. This illustrates that the conductivity fluctuations induced by oxygen vacancies not only shorten the decoherence time but also directly degrade the qubit's ability to sustain coherent gate operations. Therefore, controlling V$_O$ formation---both coordination environment and concentration---is critical for maintaining high‑fidelity quantum gates in superconducting quantum processors, especially in radiation‑prone environments.

\section{Conclusion}
\label{sec:conclusion}

In summary, by constructing amorphous Al$_2$O$_3$ models containing oxygen-vacancy defects with different coordination environments and concentrations, this work employs DFT calculations to systematically investigate the influence of V$_O$s on the electrical properties of the material. The results show that both the coordination environment and the concentration of V$_O$s play a key role in determining the electrical behavior of amorphous Al$_2$O$_3$.

Compared with the defect-free model, the 4-coordinated V$_O$ induces pronounced conductivity fluctuations, leading to a substantial reduction in the coherence time of the qubit. In contrast, 2- and 3-coordinated V$_O$s tend to introduce shallow donor states, which enhance the electrical conductivity while causing relatively smaller conductivity fluctuations. Moreover, the concentration of V$_O$s exhibits a nonlinear modulation effect on the electrical properties, and as the defect density increases, the amplitude of conductivity fluctuations grows, resulting in progressively shorter coherence times of the qubit.

Overall, this work provides a comprehensive analysis of the effects of irradiation-induced V$_O$s in the amorphous Al$_2$O$_3$ tunnel barrier of Josephson junctions on both the electrical properties of the material and the coherence of the qubit. Our findings emphasize the importance of controlling V$_O$ formation---both in terms of coordination environment and concentration---for improving the radiation tolerance and coherence performance of superconducting quantum devices. Future work could involve experimental validation of these predictions through controlled irradiation experiments and noise measurements. Additionally, multi-scale modeling that connects atomistic defect structures to device-level noise characteristics, as well as exploring alternative barrier materials or defect-engineering strategies (such as doping or passivation), may offer pathways to further enhance qubit coherence under irradiation environments.

\bibliography{a.bib}

\begin{thebibliography}{35}%
\makeatletter
\providecommand \@ifxundefined [1]{%
 \@ifx{#1\undefined}
}%
\providecommand \@ifnum [1]{%
 \ifnum #1\expandafter \@firstoftwo
 \else \expandafter \@secondoftwo
 \fi
}%
\providecommand \@ifx [1]{%
 \ifx #1\expandafter \@firstoftwo
 \else \expandafter \@secondoftwo
 \fi
}%
\providecommand \natexlab [1]{#1}%
\providecommand \enquote  [1]{``#1''}%
\providecommand \bibnamefont  [1]{#1}%
\providecommand \bibfnamefont [1]{#1}%
\providecommand \citenamefont [1]{#1}%
\providecommand \href@noop [0]{\@secondoftwo}%
\providecommand \href [0]{\begingroup \@sanitize@url \@href}%
\providecommand \@href[1]{\@@startlink{#1}\@@href}%
\providecommand \@@href[1]{\endgroup#1\@@endlink}%
\providecommand \@sanitize@url [0]{\catcode `\\12\catcode `\$12\catcode
  `\&12\catcode `\#12\catcode `\^12\catcode `\_12\catcode `\%12\relax}%
\providecommand \@@startlink[1]{}%
\providecommand \@@endlink[0]{}%
\providecommand \url  [0]{\begingroup\@sanitize@url \@url }%
\providecommand \@url [1]{\endgroup\@href {#1}{\urlprefix }}%
\providecommand \urlprefix  [0]{URL }%
\providecommand \Eprint [0]{\href }%
\providecommand \doibase [0]{https://doi.org/}%
\providecommand \selectlanguage [0]{\@gobble}%
\providecommand \bibinfo  [0]{\@secondoftwo}%
\providecommand \bibfield  [0]{\@secondoftwo}%
\providecommand \translation [1]{[#1]}%
\providecommand \BibitemOpen [0]{}%
\providecommand \bibitemStop [0]{}%
\providecommand \bibitemNoStop [0]{.\EOS\space}%
\providecommand \EOS [0]{\spacefactor3000\relax}%
\providecommand \BibitemShut  [1]{\csname bibitem#1\endcsname}%
\let\auto@bib@innerbib\@empty
\bibitem [{\citenamefont {Nielsen}\ and\ \citenamefont
  {Chuang}(2010)}]{nielsen2010quantum}%
  \BibitemOpen
  \bibfield  {author} {\bibinfo {author} {\bibfnamefont {M.~A.}\ \bibnamefont
  {Nielsen}}\ and\ \bibinfo {author} {\bibfnamefont {I.~L.}\ \bibnamefont
  {Chuang}},\ }\href@noop {} {\emph {\bibinfo {title} {Quantum computation and
  quantum information}}}\ (\bibinfo  {publisher} {Cambridge university press},\
  \bibinfo {year} {2010})\BibitemShut {NoStop}%
\bibitem [{\citenamefont {Sanders}(2025)}]{sanders2025superconducting}%
  \BibitemOpen
  \bibfield  {author} {\bibinfo {author} {\bibfnamefont {B.~C.}\ \bibnamefont
  {Sanders}},\ }\bibfield  {title} {\bibinfo {title} {Superconducting quantum
  computing beyond 100 qubits},\ }\href@noop {} {\bibfield  {journal} {\bibinfo
   {journal} {Physics}\ }\textbf {\bibinfo {volume} {18}},\ \bibinfo {pages}
  {45} (\bibinfo {year} {2025})}\BibitemShut {NoStop}%
\bibitem [{\citenamefont {Koch}\ \emph {et~al.}(2007)\citenamefont {Koch},
  \citenamefont {Yu}, \citenamefont {Gambetta}, \citenamefont {Houck},
  \citenamefont {Schuster}, \citenamefont {Majer}, \citenamefont {Blais},
  \citenamefont {Devoret}, \citenamefont {Girvin},\ and\ \citenamefont
  {Schoelkopf}}]{dielectricloss16}%
  \BibitemOpen
  \bibfield  {author} {\bibinfo {author} {\bibfnamefont {J.}~\bibnamefont
  {Koch}}, \bibinfo {author} {\bibfnamefont {T.~M.}\ \bibnamefont {Yu}},
  \bibinfo {author} {\bibfnamefont {J.}~\bibnamefont {Gambetta}}, \bibinfo
  {author} {\bibfnamefont {A.~A.}\ \bibnamefont {Houck}}, \bibinfo {author}
  {\bibfnamefont {D.~I.}\ \bibnamefont {Schuster}}, \bibinfo {author}
  {\bibfnamefont {J.}~\bibnamefont {Majer}}, \bibinfo {author} {\bibfnamefont
  {A.}~\bibnamefont {Blais}}, \bibinfo {author} {\bibfnamefont {M.~H.}\
  \bibnamefont {Devoret}}, \bibinfo {author} {\bibfnamefont {S.~M.}\
  \bibnamefont {Girvin}},\ and\ \bibinfo {author} {\bibfnamefont {R.~J.}\
  \bibnamefont {Schoelkopf}},\ }\bibfield  {title} {\bibinfo {title}
  {Charge-insensitive qubit design derived from the cooper pair box},\
  }\href@noop {} {\bibfield  {journal} {\bibinfo  {journal} {Physical Review
  A—Atomic, Molecular, and Optical Physics}\ }\textbf {\bibinfo {volume}
  {76}},\ \bibinfo {pages} {042319} (\bibinfo {year} {2007})}\BibitemShut
  {NoStop}%
\bibitem [{\citenamefont {Bennett}\ \emph {et~al.}(2009)\citenamefont
  {Bennett}, \citenamefont {Longobardi}, \citenamefont {Patel}, \citenamefont
  {Chen}, \citenamefont {Averin},\ and\ \citenamefont {Lukens}}]{rf-squid17}%
  \BibitemOpen
  \bibfield  {author} {\bibinfo {author} {\bibfnamefont {D.~A.}\ \bibnamefont
  {Bennett}}, \bibinfo {author} {\bibfnamefont {L.}~\bibnamefont {Longobardi}},
  \bibinfo {author} {\bibfnamefont {V.}~\bibnamefont {Patel}}, \bibinfo
  {author} {\bibfnamefont {W.}~\bibnamefont {Chen}}, \bibinfo {author}
  {\bibfnamefont {D.~V.}\ \bibnamefont {Averin}},\ and\ \bibinfo {author}
  {\bibfnamefont {J.~E.}\ \bibnamefont {Lukens}},\ }\bibfield  {title}
  {\bibinfo {title} {Decoherence in rf squid qubits},\ }\href@noop {}
  {\bibfield  {journal} {\bibinfo  {journal} {Quantum Information Processing}\
  }\textbf {\bibinfo {volume} {8}},\ \bibinfo {pages} {217} (\bibinfo {year}
  {2009})}\BibitemShut {NoStop}%
\bibitem [{\citenamefont {Catelani}\ \emph {et~al.}(2012)\citenamefont
  {Catelani}, \citenamefont {Nigg}, \citenamefont {Girvin}, \citenamefont
  {Schoelkopf},\ and\ \citenamefont {Glazman}}]{quasiparticle18}%
  \BibitemOpen
  \bibfield  {author} {\bibinfo {author} {\bibfnamefont {G.}~\bibnamefont
  {Catelani}}, \bibinfo {author} {\bibfnamefont {S.~E.}\ \bibnamefont {Nigg}},
  \bibinfo {author} {\bibfnamefont {S.~M.}\ \bibnamefont {Girvin}}, \bibinfo
  {author} {\bibfnamefont {R.~J.}\ \bibnamefont {Schoelkopf}},\ and\ \bibinfo
  {author} {\bibfnamefont {L.~I.}\ \bibnamefont {Glazman}},\ }\bibfield
  {title} {\bibinfo {title} {Decoherence of superconducting qubits caused by
  quasiparticle tunneling},\ }\href@noop {} {\bibfield  {journal} {\bibinfo
  {journal} {Physical Review B—Condensed Matter and Materials Physics}\
  }\textbf {\bibinfo {volume} {86}},\ \bibinfo {pages} {184514} (\bibinfo
  {year} {2012})}\BibitemShut {NoStop}%
\bibitem [{\citenamefont {Pan}\ \emph {et~al.}(2022)\citenamefont {Pan},
  \citenamefont {Zhou}, \citenamefont {Yuan}, \citenamefont {Nie},
  \citenamefont {Wei}, \citenamefont {Zhang}, \citenamefont {Li}, \citenamefont
  {Liu}, \citenamefont {Jiang}, \citenamefont {Catelani} \emph
  {et~al.}}]{quasiparticles-chare19}%
  \BibitemOpen
  \bibfield  {author} {\bibinfo {author} {\bibfnamefont {X.}~\bibnamefont
  {Pan}}, \bibinfo {author} {\bibfnamefont {Y.}~\bibnamefont {Zhou}}, \bibinfo
  {author} {\bibfnamefont {H.}~\bibnamefont {Yuan}}, \bibinfo {author}
  {\bibfnamefont {L.}~\bibnamefont {Nie}}, \bibinfo {author} {\bibfnamefont
  {W.}~\bibnamefont {Wei}}, \bibinfo {author} {\bibfnamefont {L.}~\bibnamefont
  {Zhang}}, \bibinfo {author} {\bibfnamefont {J.}~\bibnamefont {Li}}, \bibinfo
  {author} {\bibfnamefont {S.}~\bibnamefont {Liu}}, \bibinfo {author}
  {\bibfnamefont {Z.~H.}\ \bibnamefont {Jiang}}, \bibinfo {author}
  {\bibfnamefont {G.}~\bibnamefont {Catelani}}, \emph {et~al.},\ }\bibfield
  {title} {\bibinfo {title} {Engineering superconducting qubits to reduce
  quasiparticles and charge noise},\ }\href@noop {} {\bibfield  {journal}
  {\bibinfo  {journal} {Nature Communications}\ }\textbf {\bibinfo {volume}
  {13}},\ \bibinfo {pages} {7196} (\bibinfo {year} {2022})}\BibitemShut
  {NoStop}%
\bibitem [{\citenamefont {Shalibo}\ \emph {et~al.}(2010)\citenamefont
  {Shalibo}, \citenamefont {Rofe}, \citenamefont {Shwa}, \citenamefont
  {Zeides}, \citenamefont {Neeley}, \citenamefont {Martinis},\ and\
  \citenamefont {Katz}}]{tls-lifetime-and-coherence20}%
  \BibitemOpen
  \bibfield  {author} {\bibinfo {author} {\bibfnamefont {Y.}~\bibnamefont
  {Shalibo}}, \bibinfo {author} {\bibfnamefont {Y.}~\bibnamefont {Rofe}},
  \bibinfo {author} {\bibfnamefont {D.}~\bibnamefont {Shwa}}, \bibinfo {author}
  {\bibfnamefont {F.}~\bibnamefont {Zeides}}, \bibinfo {author} {\bibfnamefont
  {M.}~\bibnamefont {Neeley}}, \bibinfo {author} {\bibfnamefont {J.~M.}\
  \bibnamefont {Martinis}},\ and\ \bibinfo {author} {\bibfnamefont
  {N.}~\bibnamefont {Katz}},\ }\bibfield  {title} {\bibinfo {title} {Lifetime
  and coherence of two-level defects in a josephson junction},\ }\href@noop {}
  {\bibfield  {journal} {\bibinfo  {journal} {Physical review letters}\
  }\textbf {\bibinfo {volume} {105}},\ \bibinfo {pages} {177001} (\bibinfo
  {year} {2010})}\BibitemShut {NoStop}%
\bibitem [{\citenamefont {Choi}(2010)}]{fulx-noise21}%
  \BibitemOpen
  \bibfield  {author} {\bibinfo {author} {\bibfnamefont {S.}~\bibnamefont
  {Choi}},\ }\bibfield  {title} {\bibinfo {title} {Localization of
  metal-induced gap states at the metal-insulator interface: Origin of flux
  noise},\ }\href@noop {} {\  (\bibinfo {year} {2010})}\BibitemShut {NoStop}%
\bibitem [{\citenamefont {M{\"u}ller}\ \emph {et~al.}(2009)\citenamefont
  {M{\"u}ller}, \citenamefont {Shnirman},\ and\ \citenamefont
  {Makhlin}}]{tls-relaxtion22}%
  \BibitemOpen
  \bibfield  {author} {\bibinfo {author} {\bibfnamefont {C.}~\bibnamefont
  {M{\"u}ller}}, \bibinfo {author} {\bibfnamefont {A.}~\bibnamefont
  {Shnirman}},\ and\ \bibinfo {author} {\bibfnamefont {Y.}~\bibnamefont
  {Makhlin}},\ }\bibfield  {title} {\bibinfo {title} {Relaxation of josephson
  qubits due to strong coupling to two-level systems},\ }\href@noop {}
  {\bibfield  {journal} {\bibinfo  {journal} {Physical Review B—Condensed
  Matter and Materials Physics}\ }\textbf {\bibinfo {volume} {80}},\ \bibinfo
  {pages} {134517} (\bibinfo {year} {2009})}\BibitemShut {NoStop}%
\bibitem [{\citenamefont {Holder}\ \emph {et~al.}(2013)\citenamefont {Holder},
  \citenamefont {Osborn}, \citenamefont {Lobb},\ and\ \citenamefont
  {Musgrave}}]{holder2013bulk25}%
  \BibitemOpen
  \bibfield  {author} {\bibinfo {author} {\bibfnamefont {A.~M.}\ \bibnamefont
  {Holder}}, \bibinfo {author} {\bibfnamefont {K.~D.}\ \bibnamefont {Osborn}},
  \bibinfo {author} {\bibfnamefont {C.}~\bibnamefont {Lobb}},\ and\ \bibinfo
  {author} {\bibfnamefont {C.~B.}\ \bibnamefont {Musgrave}},\ }\bibfield
  {title} {\bibinfo {title} {Bulk and surface tunneling hydrogen defects in
  alumina},\ }\href@noop {} {\bibfield  {journal} {\bibinfo  {journal}
  {Physical review letters}\ }\textbf {\bibinfo {volume} {111}},\ \bibinfo
  {pages} {065901} (\bibinfo {year} {2013})}\BibitemShut {NoStop}%
\bibitem [{\citenamefont {Gordon}\ \emph {et~al.}(2014)\citenamefont {Gordon},
  \citenamefont {Abu-Farsakh}, \citenamefont {Janotti},\ and\ \citenamefont
  {Van~de Walle}}]{O-H...Hhydrogen}%
  \BibitemOpen
  \bibfield  {author} {\bibinfo {author} {\bibfnamefont {L.}~\bibnamefont
  {Gordon}}, \bibinfo {author} {\bibfnamefont {H.}~\bibnamefont {Abu-Farsakh}},
  \bibinfo {author} {\bibfnamefont {A.}~\bibnamefont {Janotti}},\ and\ \bibinfo
  {author} {\bibfnamefont {C.~G.}\ \bibnamefont {Van~de Walle}},\ }\bibfield
  {title} {\bibinfo {title} {Hydrogen bonds in al2o3 as dissipative two-level
  systems in superconducting qubits},\ }\href@noop {} {\bibfield  {journal}
  {\bibinfo  {journal} {Scientific reports}\ }\textbf {\bibinfo {volume} {4}},\
  \bibinfo {pages} {7590} (\bibinfo {year} {2014})}\BibitemShut {NoStop}%
\bibitem [{\citenamefont {Bafia}\ \emph {et~al.}(2024)\citenamefont {Bafia},
  \citenamefont {Murthy}, \citenamefont {Grassellino},\ and\ \citenamefont
  {Romanenko}}]{oxygen37}%
  \BibitemOpen
  \bibfield  {author} {\bibinfo {author} {\bibfnamefont {D.}~\bibnamefont
  {Bafia}}, \bibinfo {author} {\bibfnamefont {A.}~\bibnamefont {Murthy}},
  \bibinfo {author} {\bibfnamefont {A.}~\bibnamefont {Grassellino}},\ and\
  \bibinfo {author} {\bibfnamefont {A.}~\bibnamefont {Romanenko}},\ }\bibfield
  {title} {\bibinfo {title} {Oxygen vacancies in niobium pentoxide as a source
  of two-level system losses in superconducting niobium},\ }\href@noop {}
  {\bibfield  {journal} {\bibinfo  {journal} {Physical Review Applied}\
  }\textbf {\bibinfo {volume} {22}},\ \bibinfo {pages} {024035} (\bibinfo
  {year} {2024})}\BibitemShut {NoStop}%
\bibitem [{\citenamefont {Guo}\ \emph {et~al.}(2016)\citenamefont {Guo},
  \citenamefont {Ambrosio},\ and\ \citenamefont
  {Pasquarello}}]{guo2016oxygen38}%
  \BibitemOpen
  \bibfield  {author} {\bibinfo {author} {\bibfnamefont {Z.}~\bibnamefont
  {Guo}}, \bibinfo {author} {\bibfnamefont {F.}~\bibnamefont {Ambrosio}},\ and\
  \bibinfo {author} {\bibfnamefont {A.}~\bibnamefont {Pasquarello}},\
  }\bibfield  {title} {\bibinfo {title} {Oxygen defects in amorphous al2o3: A
  hybrid functional study},\ }\href@noop {} {\bibfield  {journal} {\bibinfo
  {journal} {Applied physics letters}\ }\textbf {\bibinfo {volume} {109}}
  (\bibinfo {year} {2016})}\BibitemShut {NoStop}%
\bibitem [{\citenamefont {Qiu}\ \emph {et~al.}(2024)\citenamefont {Qiu},
  \citenamefont {Wang}, \citenamefont {Sun}, \citenamefont {Han},\ and\
  \citenamefont {Shan}}]{qiu2024manipulation39}%
  \BibitemOpen
  \bibfield  {author} {\bibinfo {author} {\bibfnamefont {J.}~\bibnamefont
  {Qiu}}, \bibinfo {author} {\bibfnamefont {S.}~\bibnamefont {Wang}}, \bibinfo
  {author} {\bibfnamefont {H.}~\bibnamefont {Sun}}, \bibinfo {author}
  {\bibfnamefont {C.}~\bibnamefont {Han}},\ and\ \bibinfo {author}
  {\bibfnamefont {Z.}~\bibnamefont {Shan}},\ }\bibfield  {title} {\bibinfo
  {title} {Manipulation of electrical performance in al-based josephson
  junctions via oxygen vacancies in barrier},\ }\href@noop {} {\bibfield
  {journal} {\bibinfo  {journal} {Journal of Materials Chemistry C}\ }\textbf
  {\bibinfo {volume} {12}},\ \bibinfo {pages} {19063} (\bibinfo {year}
  {2024})}\BibitemShut {NoStop}%
\bibitem [{\citenamefont {Ji}(2008)}]{envrionment-entagle}%
  \BibitemOpen
  \bibfield  {author} {\bibinfo {author} {\bibfnamefont {Y.-H.}\ \bibnamefont
  {Ji}},\ }\bibfield  {title} {\bibinfo {title} {Investigation of decoherence
  of superconducting charge qubit entangled with the environment},\ }\href@noop
  {} {\bibfield  {journal} {\bibinfo  {journal} {International Journal of
  Theoretical Physics}\ }\textbf {\bibinfo {volume} {47}},\ \bibinfo {pages}
  {2363} (\bibinfo {year} {2008})}\BibitemShut {NoStop}%
\bibitem [{\citenamefont {Koppinen}\ \emph {et~al.}(2007)\citenamefont
  {Koppinen}, \citenamefont {V{\"a}ist{\"o}},\ and\ \citenamefont
  {Maasilta}}]{annealing67}%
  \BibitemOpen
  \bibfield  {author} {\bibinfo {author} {\bibfnamefont {P.}~\bibnamefont
  {Koppinen}}, \bibinfo {author} {\bibfnamefont {L.}~\bibnamefont
  {V{\"a}ist{\"o}}},\ and\ \bibinfo {author} {\bibfnamefont {I.}~\bibnamefont
  {Maasilta}},\ }\bibfield  {title} {\bibinfo {title} {Complete stabilization
  and improvement of the characteristics of tunnel junctions by thermal
  annealing},\ }\href@noop {} {\bibfield  {journal} {\bibinfo  {journal}
  {Applied physics letters}\ }\textbf {\bibinfo {volume} {90}} (\bibinfo {year}
  {2007})}\BibitemShut {NoStop}%
\bibitem [{\citenamefont {Korshakov}\ \emph {et~al.}(2024)\citenamefont
  {Korshakov}, \citenamefont {Moskalev}, \citenamefont {Soloveva},
  \citenamefont {Moskaleva}, \citenamefont {Lotkov}, \citenamefont {Ibragimov},
  \citenamefont {Androschuk}, \citenamefont {Ryzhikov}, \citenamefont
  {Panfilov},\ and\ \citenamefont {Rodionov}}]{anneling}%
  \BibitemOpen
  \bibfield  {author} {\bibinfo {author} {\bibfnamefont {N.~D.}\ \bibnamefont
  {Korshakov}}, \bibinfo {author} {\bibfnamefont {D.~O.}\ \bibnamefont
  {Moskalev}}, \bibinfo {author} {\bibfnamefont {A.~A.}\ \bibnamefont
  {Soloveva}}, \bibinfo {author} {\bibfnamefont {D.~A.}\ \bibnamefont
  {Moskaleva}}, \bibinfo {author} {\bibfnamefont {E.~S.}\ \bibnamefont
  {Lotkov}}, \bibinfo {author} {\bibfnamefont {A.~R.}\ \bibnamefont
  {Ibragimov}}, \bibinfo {author} {\bibfnamefont {M.~V.}\ \bibnamefont
  {Androschuk}}, \bibinfo {author} {\bibfnamefont {I.~A.}\ \bibnamefont
  {Ryzhikov}}, \bibinfo {author} {\bibfnamefont {Y.~V.}\ \bibnamefont
  {Panfilov}},\ and\ \bibinfo {author} {\bibfnamefont {I.~A.}\ \bibnamefont
  {Rodionov}},\ }\bibfield  {title} {\bibinfo {title} {Aluminum josephson
  junction microstructure and electrical properties modified by thermal
  annealing},\ }\href@noop {} {\bibfield  {journal} {\bibinfo  {journal}
  {Scientific Reports}\ }\textbf {\bibinfo {volume} {14}},\ \bibinfo {pages}
  {26066} (\bibinfo {year} {2024})}\BibitemShut {NoStop}%
\bibitem [{\citenamefont {M{\"u}ller}\ \emph {et~al.}(2019)\citenamefont
  {M{\"u}ller}, \citenamefont {Cole},\ and\ \citenamefont
  {Lisenfeld}}]{muller2019towards78}%
  \BibitemOpen
  \bibfield  {author} {\bibinfo {author} {\bibfnamefont {C.}~\bibnamefont
  {M{\"u}ller}}, \bibinfo {author} {\bibfnamefont {J.~H.}\ \bibnamefont
  {Cole}},\ and\ \bibinfo {author} {\bibfnamefont {J.}~\bibnamefont
  {Lisenfeld}},\ }\bibfield  {title} {\bibinfo {title} {Towards understanding
  two-level-systems in amorphous solids: insights from quantum circuits},\
  }\href@noop {} {\bibfield  {journal} {\bibinfo  {journal} {Reports on
  Progress in Physics}\ }\textbf {\bibinfo {volume} {82}},\ \bibinfo {pages}
  {124501} (\bibinfo {year} {2019})}\BibitemShut {NoStop}%
\bibitem [{\citenamefont {Ding}(2021)}]{ding2021damage49}%
  \BibitemOpen
  \bibfield  {author} {\bibinfo {author} {\bibfnamefont {M.}~\bibnamefont
  {Ding}},\ }\bibfield  {title} {\bibinfo {title} {Damage effect of ald-al2o3
  based metal-oxide-semiconductor structures under gamma-ray irradiation},\
  }\href@noop {} {\bibfield  {journal} {\bibinfo  {journal} {Micromachines}\
  }\textbf {\bibinfo {volume} {12}},\ \bibinfo {pages} {661} (\bibinfo {year}
  {2021})}\BibitemShut {NoStop}%
\bibitem [{\citenamefont {Ding}\ \emph {et~al.}(2014)\citenamefont {Ding},
  \citenamefont {Cheng}, \citenamefont {Liu},\ and\ \citenamefont
  {Li}}]{ding2014total50}%
  \BibitemOpen
  \bibfield  {author} {\bibinfo {author} {\bibfnamefont {M.}~\bibnamefont
  {Ding}}, \bibinfo {author} {\bibfnamefont {Y.}~\bibnamefont {Cheng}},
  \bibinfo {author} {\bibfnamefont {X.}~\bibnamefont {Liu}},\ and\ \bibinfo
  {author} {\bibfnamefont {X.}~\bibnamefont {Li}},\ }\bibfield  {title}
  {\bibinfo {title} {Total dose response of hafnium oxide based
  metal-oxide-semiconductor structure under gamma-ray irradiation},\
  }\href@noop {} {\bibfield  {journal} {\bibinfo  {journal} {IEEE Transactions
  on Dielectrics and Electrical Insulation}\ }\textbf {\bibinfo {volume}
  {21}},\ \bibinfo {pages} {1792} (\bibinfo {year} {2014})}\BibitemShut
  {NoStop}%
\bibitem [{\citenamefont {Zhu}\ \emph {et~al.}(2018)\citenamefont {Zhu},
  \citenamefont {Zheng}, \citenamefont {Li}, \citenamefont {Li}, \citenamefont
  {Zhang}, \citenamefont {Li}, \citenamefont {Gao}, \citenamefont {Yang},
  \citenamefont {Cui}, \citenamefont {Liang} \emph {et~al.}}]{zhu2018total51}%
  \BibitemOpen
  \bibfield  {author} {\bibinfo {author} {\bibfnamefont {H.}~\bibnamefont
  {Zhu}}, \bibinfo {author} {\bibfnamefont {Z.}~\bibnamefont {Zheng}}, \bibinfo
  {author} {\bibfnamefont {B.}~\bibnamefont {Li}}, \bibinfo {author}
  {\bibfnamefont {B.}~\bibnamefont {Li}}, \bibinfo {author} {\bibfnamefont
  {G.}~\bibnamefont {Zhang}}, \bibinfo {author} {\bibfnamefont
  {D.}~\bibnamefont {Li}}, \bibinfo {author} {\bibfnamefont {J.}~\bibnamefont
  {Gao}}, \bibinfo {author} {\bibfnamefont {L.}~\bibnamefont {Yang}}, \bibinfo
  {author} {\bibfnamefont {Y.}~\bibnamefont {Cui}}, \bibinfo {author}
  {\bibfnamefont {C.}~\bibnamefont {Liang}}, \emph {et~al.},\ }\bibfield
  {title} {\bibinfo {title} {Total dose effect of al2o3-based
  metal--oxide--semiconductor structures and its mechanism under gamma-ray
  irradiation},\ }\href@noop {} {\bibfield  {journal} {\bibinfo  {journal}
  {Semiconductor Science and Technology}\ }\textbf {\bibinfo {volume} {33}},\
  \bibinfo {pages} {115010} (\bibinfo {year} {2018})}\BibitemShut {NoStop}%
\bibitem [{\citenamefont {Kresse}\ and\ \citenamefont
  {Hafner}(1993)}]{VASP_kresse1993ab}%
  \BibitemOpen
  \bibfield  {author} {\bibinfo {author} {\bibfnamefont {G.}~\bibnamefont
  {Kresse}}\ and\ \bibinfo {author} {\bibfnamefont {J.}~\bibnamefont
  {Hafner}},\ }\bibfield  {title} {\bibinfo {title} {Ab initio molecular
  dynamics for liquid metals},\ }\href@noop {} {\bibfield  {journal} {\bibinfo
  {journal} {Physical review B}\ }\textbf {\bibinfo {volume} {47}},\ \bibinfo
  {pages} {558} (\bibinfo {year} {1993})}\BibitemShut {NoStop}%
\bibitem [{\citenamefont {Kresse}\ and\ \citenamefont
  {Furthm{\"u}ller}(1996)}]{Vasp_kresse1996efficient}%
  \BibitemOpen
  \bibfield  {author} {\bibinfo {author} {\bibfnamefont {G.}~\bibnamefont
  {Kresse}}\ and\ \bibinfo {author} {\bibfnamefont {J.}~\bibnamefont
  {Furthm{\"u}ller}},\ }\bibfield  {title} {\bibinfo {title} {Efficient
  iterative schemes for ab initio total-energy calculations using a plane-wave
  basis set},\ }\href@noop {} {\bibfield  {journal} {\bibinfo  {journal}
  {Physical review B}\ }\textbf {\bibinfo {volume} {54}},\ \bibinfo {pages}
  {11169} (\bibinfo {year} {1996})}\BibitemShut {NoStop}%
\bibitem [{\citenamefont {Bl{\"o}chl}(1994)}]{PAW_blochl1994projector}%
  \BibitemOpen
  \bibfield  {author} {\bibinfo {author} {\bibfnamefont {P.~E.}\ \bibnamefont
  {Bl{\"o}chl}},\ }\bibfield  {title} {\bibinfo {title} {Projector
  augmented-wave method},\ }\href@noop {} {\bibfield  {journal} {\bibinfo
  {journal} {Physical review B}\ }\textbf {\bibinfo {volume} {50}},\ \bibinfo
  {pages} {17953} (\bibinfo {year} {1994})}\BibitemShut {NoStop}%
\bibitem [{\citenamefont {Kresse}\ and\ \citenamefont
  {Joubert}(1999)}]{PAW_kresse1999ultrasoft}%
  \BibitemOpen
  \bibfield  {author} {\bibinfo {author} {\bibfnamefont {G.}~\bibnamefont
  {Kresse}}\ and\ \bibinfo {author} {\bibfnamefont {D.}~\bibnamefont
  {Joubert}},\ }\bibfield  {title} {\bibinfo {title} {From ultrasoft
  pseudopotentials to the projector augmented-wave method},\ }\href@noop {}
  {\bibfield  {journal} {\bibinfo  {journal} {Physical review b}\ }\textbf
  {\bibinfo {volume} {59}},\ \bibinfo {pages} {1758} (\bibinfo {year}
  {1999})}\BibitemShut {NoStop}%
\bibitem [{\citenamefont {Sun}\ \emph {et~al.}(2015)\citenamefont {Sun},
  \citenamefont {Ruzsinszky},\ and\ \citenamefont
  {Perdew}}]{Scan_sun2015strongly}%
  \BibitemOpen
  \bibfield  {author} {\bibinfo {author} {\bibfnamefont {J.}~\bibnamefont
  {Sun}}, \bibinfo {author} {\bibfnamefont {A.}~\bibnamefont {Ruzsinszky}},\
  and\ \bibinfo {author} {\bibfnamefont {J.~P.}\ \bibnamefont {Perdew}},\
  }\bibfield  {title} {\bibinfo {title} {Strongly constrained and appropriately
  normed semilocal density functional},\ }\href@noop {} {\bibfield  {journal}
  {\bibinfo  {journal} {Physical review letters}\ }\textbf {\bibinfo {volume}
  {115}},\ \bibinfo {pages} {036402} (\bibinfo {year} {2015})}\BibitemShut
  {NoStop}%
\bibitem [{\citenamefont {Heyd}\ \emph {et~al.}(2003)\citenamefont {Heyd},
  \citenamefont {Scuseria},\ and\ \citenamefont
  {Ernzerhof}}]{Hse_heyd2003hybrid}%
  \BibitemOpen
  \bibfield  {author} {\bibinfo {author} {\bibfnamefont {J.}~\bibnamefont
  {Heyd}}, \bibinfo {author} {\bibfnamefont {G.~E.}\ \bibnamefont {Scuseria}},\
  and\ \bibinfo {author} {\bibfnamefont {M.}~\bibnamefont {Ernzerhof}},\
  }\bibfield  {title} {\bibinfo {title} {Hybrid functionals based on a screened
  coulomb potential},\ }\href@noop {} {\bibfield  {journal} {\bibinfo
  {journal} {The Journal of chemical physics}\ }\textbf {\bibinfo {volume}
  {118}},\ \bibinfo {pages} {8207} (\bibinfo {year} {2003})}\BibitemShut
  {NoStop}%
\bibitem [{\citenamefont {Ge}\ and\ \citenamefont
  {Ernzerhof}(2006)}]{Hse_ge2006erratum}%
  \BibitemOpen
  \bibfield  {author} {\bibinfo {author} {\bibfnamefont {H.~J.~S.}\
  \bibnamefont {Ge}}\ and\ \bibinfo {author} {\bibfnamefont {M.}~\bibnamefont
  {Ernzerhof}},\ }\bibfield  {title} {\bibinfo {title} {Erratum:“hybrid
  functionals based on a screened coulomb potential”[j. chem. phys. 118, 8207
  (2003)]},\ }\href@noop {} {\bibfield  {journal} {\bibinfo  {journal} {J.
  Chem. Phys}\ }\textbf {\bibinfo {volume} {124}},\ \bibinfo {pages} {219906}
  (\bibinfo {year} {2006})}\BibitemShut {NoStop}%
\bibitem [{\citenamefont {Filatova}\ and\ \citenamefont
  {Konashuk}(2015)}]{band_gap_7-0.1}%
  \BibitemOpen
  \bibfield  {author} {\bibinfo {author} {\bibfnamefont {E.~O.}\ \bibnamefont
  {Filatova}}\ and\ \bibinfo {author} {\bibfnamefont {A.~S.}\ \bibnamefont
  {Konashuk}},\ }\bibfield  {title} {\bibinfo {title} {Interpretation of the
  changing the band gap of al2o3 depending on its crystalline form: connection
  with different local symmetries},\ }\href@noop {} {\bibfield  {journal}
  {\bibinfo  {journal} {The Journal of Physical Chemistry C}\ }\textbf
  {\bibinfo {volume} {119}},\ \bibinfo {pages} {20755} (\bibinfo {year}
  {2015})}\BibitemShut {NoStop}%
\bibitem [{\citenamefont {Pugliese}\ \emph {et~al.}(2022)\citenamefont
  {Pugliese}, \citenamefont {Shyam}, \citenamefont {Repa}, \citenamefont
  {Nguyen}, \citenamefont {Mehta}, \citenamefont {Webb~III}, \citenamefont
  {Fredin},\ and\ \citenamefont {Strandwitz}}]{amphous_band_gap_6-7}%
  \BibitemOpen
  \bibfield  {author} {\bibinfo {author} {\bibfnamefont {A.}~\bibnamefont
  {Pugliese}}, \bibinfo {author} {\bibfnamefont {B.}~\bibnamefont {Shyam}},
  \bibinfo {author} {\bibfnamefont {G.~M.}\ \bibnamefont {Repa}}, \bibinfo
  {author} {\bibfnamefont {A.~H.}\ \bibnamefont {Nguyen}}, \bibinfo {author}
  {\bibfnamefont {A.}~\bibnamefont {Mehta}}, \bibinfo {author} {\bibfnamefont
  {E.~B.}\ \bibnamefont {Webb~III}}, \bibinfo {author} {\bibfnamefont {L.~A.}\
  \bibnamefont {Fredin}},\ and\ \bibinfo {author} {\bibfnamefont {N.~C.}\
  \bibnamefont {Strandwitz}},\ }\bibfield  {title} {\bibinfo {title}
  {Atomic-layer-deposited aluminum oxide thin films probed with x-ray
  scattering and compared to molecular dynamics and density functional theory
  models},\ }\href@noop {} {\bibfield  {journal} {\bibinfo  {journal} {ACS
  omega}\ }\textbf {\bibinfo {volume} {7}},\ \bibinfo {pages} {41033} (\bibinfo
  {year} {2022})}\BibitemShut {NoStop}%
\bibitem [{\citenamefont {Gutierrez}\ and\ \citenamefont
  {Johansson}(2002)}]{gutierrez2002molecular}%
  \BibitemOpen
  \bibfield  {author} {\bibinfo {author} {\bibfnamefont {G.}~\bibnamefont
  {Gutierrez}}\ and\ \bibinfo {author} {\bibfnamefont {B.}~\bibnamefont
  {Johansson}},\ }\bibfield  {title} {\bibinfo {title} {Molecular dynamics
  study of structural properties of amorphous al 2 o 3},\ }\href@noop {}
  {\bibfield  {journal} {\bibinfo  {journal} {Physical review B}\ }\textbf
  {\bibinfo {volume} {65}},\ \bibinfo {pages} {104202} (\bibinfo {year}
  {2002})}\BibitemShut {NoStop}%
\bibitem [{\citenamefont {Van~Hoang}\ and\ \citenamefont
  {Oh}(2004)}]{structure}%
  \BibitemOpen
  \bibfield  {author} {\bibinfo {author} {\bibfnamefont {V.}~\bibnamefont
  {Van~Hoang}}\ and\ \bibinfo {author} {\bibfnamefont {S.~K.}\ \bibnamefont
  {Oh}},\ }\bibfield  {title} {\bibinfo {title} {Simulation of structural
  properties and structural transformation of amorphous al2o3},\ }\href@noop {}
  {\bibfield  {journal} {\bibinfo  {journal} {Physica B: Condensed Matter}\
  }\textbf {\bibinfo {volume} {352}},\ \bibinfo {pages} {73} (\bibinfo {year}
  {2004})}\BibitemShut {NoStop}%
\bibitem [{\citenamefont {Lamparter}\ and\ \citenamefont
  {Kniep}(1997)}]{lamparter1997structure}%
  \BibitemOpen
  \bibfield  {author} {\bibinfo {author} {\bibfnamefont {P.}~\bibnamefont
  {Lamparter}}\ and\ \bibinfo {author} {\bibfnamefont {R.}~\bibnamefont
  {Kniep}},\ }\bibfield  {title} {\bibinfo {title} {Structure of amorphous
  al2o3},\ }\href@noop {} {\bibfield  {journal} {\bibinfo  {journal} {Physica
  B: Condensed Matter}\ }\textbf {\bibinfo {volume} {234}},\ \bibinfo {pages}
  {405} (\bibinfo {year} {1997})}\BibitemShut {NoStop}%
\bibitem [{\citenamefont {Oka}\ \emph {et~al.}(1979)\citenamefont {Oka},
  \citenamefont {Takahashi}, \citenamefont {Okada},\ and\ \citenamefont
  {Iwai}}]{oka1979structural}%
  \BibitemOpen
  \bibfield  {author} {\bibinfo {author} {\bibfnamefont {Y.}~\bibnamefont
  {Oka}}, \bibinfo {author} {\bibfnamefont {T.}~\bibnamefont {Takahashi}},
  \bibinfo {author} {\bibfnamefont {K.}~\bibnamefont {Okada}},\ and\ \bibinfo
  {author} {\bibfnamefont {S.-i.}\ \bibnamefont {Iwai}},\ }\bibfield  {title}
  {\bibinfo {title} {Structural analysis of anodic alumina films},\ }\href@noop
  {} {\bibfield  {journal} {\bibinfo  {journal} {Journal of Non-Crystalline
  Solids}\ }\textbf {\bibinfo {volume} {30}},\ \bibinfo {pages} {349} (\bibinfo
  {year} {1979})}\BibitemShut {NoStop}%
\bibitem [{\citenamefont {Van~Harlingen}\ \emph {et~al.}(2004)\citenamefont
  {Van~Harlingen}, \citenamefont {Robertson}, \citenamefont {Plourde},
  \citenamefont {Reichardt}, \citenamefont {Crane},\ and\ \citenamefont
  {Clarke}}]{vandecoherence24}%
  \BibitemOpen
  \bibfield  {author} {\bibinfo {author} {\bibfnamefont {D.}~\bibnamefont
  {Van~Harlingen}}, \bibinfo {author} {\bibfnamefont {T.}~\bibnamefont
  {Robertson}}, \bibinfo {author} {\bibfnamefont {B.}~\bibnamefont {Plourde}},
  \bibinfo {author} {\bibfnamefont {P.}~\bibnamefont {Reichardt}}, \bibinfo
  {author} {\bibfnamefont {T.}~\bibnamefont {Crane}},\ and\ \bibinfo {author}
  {\bibfnamefont {J.}~\bibnamefont {Clarke}},\ }\bibfield  {title} {\bibinfo
  {title} {Decoherence in josephson-junction qubits due to critical-current
  fluctuations},\ }\href@noop {} {\bibfield  {journal} {\bibinfo  {journal}
  {Physical Review B—Condensed Matter and Materials Physics}\ }\textbf
  {\bibinfo {volume} {70}},\ \bibinfo {pages} {064517} (\bibinfo {year}
  {2004})}\BibitemShut {NoStop}%
\end{thebibliography}%

\end{document}